\documentclass[preprintnumbers,aps,prd,longbibliography,nofootinbib,nobibnotes,amsmath,amssymb]{revtex4}
\usepackage{graphicx,color,natbib}
\usepackage{amsmath,amssymb,amsfonts}
\usepackage{epsfig,epsf}
\usepackage{dcolumn}
\usepackage{float}
\usepackage{bm}
\usepackage{subfig}
\usepackage{caption}
\input amssym.def
\input amssym.tex

\usepackage[colorlinks=true, linkcolor=blue, bookmarks=true]{hyperref}

\begin{document}
\title{Probing Phase Structure of Strongly Coupled Matter with Holographic Entanglement Measures}
\author{M. Asadi\footnote{$\rm{m}_{-}$asadi@ipm.ir}}
\affiliation{School of Particles and Accelerators, Institute for Research in Fundamental Sciences (IPM), P.O.Box 19395-5531, Tehran, Iran}
\author{B. Amrahi\footnote{$\rm{b}_{-}$amrahi@ipm.ir}}
\affiliation{School of Particles and Accelerators, Institute for Research in Fundamental Sciences (IPM), P.O.Box 19395-5531, Tehran, Iran}
\author{H. Eshaghi-Kenari\footnote{h.e.kenari@iauamol.ac.ir}}
\affiliation{Faculty of Sciences, Islamic Azad University-Ayatollah Amoli branch, Amol, P.O. Box 678, Iran}
\begin{abstract}
We study the holographic entanglement measures such as the holographic mutual information, HMI, and the holographic entanglement of purification, EoP, in a holographic QCD model at finite temperature and zero chemical potential. This model can realize various types of phase transitions including crossover, first order and second order phase transitions. We use the HMI and EoP to probe the phase structure of this model and we find that at the critical temperature they can characterize the phase structure of the model. Moreover we obtain the critical exponent using the HMI and EoP.
\end{abstract}
\maketitle

\tableofcontents
\section{Introduction}\label{sec:intro}
The gauge/gravity duality, which relates a $d$-dimensional quantum field theory (QFT) to some classical gravitational theory in ($d+1$)-dimensions, has been utilized to study various areas of physics such as condensed matter, quantum chromodynamics (QCD) and quantum information theory \cite{Maldacena:1997re,Witten:1998qj,Natsuume:2014sfa,Hartnoll:2009sz,Camilo:2016kxq,CasalderreySolana:2011us}. This duality provides a powerful tool to investigate various phenomena in the strongly coupled QFT using their corresponding gravity duals. Studying some phenomena and quantities from a field theory point of view may be difficult while this is relatively simple to study them on the gravity side. For example the confinement-deconfinement phase transition of the QCD corresponds to the Hawking-Page phase transition on the gravity side which is the transition through black hole and thermal gas backgrounds \cite{Hawking:1982dh,Gursoy:2008bu}.

In this paper we consider a holographic QCD model which is proposed in \cite{Gubser:2008ny,Gubser:2008yx} intending to mimic the equation of state of the real QCD. As we know, QCD is strongly coupled at low energy and hence it is difficult to study it using perturbation techniques. Surprisingly, the gauge/gravity duality is extended to the more general cases such as the non-conformal field theory and hence one can study the QCD in the strongly coupled region by studying its dual gravitational theory \cite{Attems:2016ugt,Pang:2015lka,Rahimi:2016bbv,Ali-Akbari:2021zsm,Taylor:2017zzo,Lezgi:2020bkc,Asadi:2021nbd,Asadi:2020gzl}. It is interesting to seek the holographic models which are dual to non-conformal field theories, such as QCD. One class of these models are top-down models which are directly constructed from string theory. One can find some examples of these models in \cite{Babington:2003vm,Kruczenski:2003uq,Kruczenski:2003be,Kobayashi:2006sb,Sakai:2004cn,Sakai:2005yt,Elander:2020rgv}. Another class of these models are bottom-up models such as the hard-wall model and the soft-wall model in which the gravitational theory are phenomenologically fixed to be in agreement with the lattice QCD data \cite{Erlich:2005qh,Karch:2006pv,Cai:2022omk}. Our considered holographic QCD model contains a dilaton field whose self-interaction potential is parameterized by some constants, and by choosing appropriate values for them the system possesses a crossover phase transition at a critical temperature and it is clearly seen that the thermodynamical properties obtained from this model are in complete agreement with the lattice QCD results \cite{Borsanyi:2012cr}. Moreover, by choosing other suitable values for the parameters in the dilaton potential, this model can also exhibit the first and second order phase transitions. Therefore this is a good model to study various phase structures of strongly coupled field systems. This model utilized for studying the phase structure of a system using the hydrodynamic transport coefficients \cite{Janik:2015waa}, quasinormal modes \cite{Janik:2016btb}, entanglement entropy \cite{Zhang:2016rcm} and complexity \cite{Zhang:2017nth}.

There are various physical quantities in the context of quantum information theory which we can study with the help of their corresponding gravity duals. One of them is the entanglement entropy (EE), a significant non-local quantity, which measures the quantum entanglement between subsystem $A$ and its complement $\bar{A}$ for a given pure state. The EE suffers from short-distance divergence which satisfies an area-law and hence it is a scheme-dependent quantity in the UV limit \cite{Bombelli:1986rw,Srednicki:1993im}. It is difficult to compute the EE in QFT. Fortunately, EE has a simple holographic dual in which the EE of a subregion $A$ in the boundary field theory corresponds to the area of minimal surface extended in the bulk whose boundary coincides with the boundary of subregion $A$ \cite{Ryu:2006bv,Ryu:2006ef}. Based on this prescription, many studies have been done in the literature to better understand the EE \cite{Casini:2011kv,Myers:2012ed,Lokhande:2017jik,Rahimi:2018ica,Fischler:2012ca,Ben-Ami:2014gsa,Pang:2014tpa,Kundu:2016dyk,Ebrahim:2020qif,Buniy:2005au,Dudal:2016joz,Dudal:2018ztm,Arefeva:2020uec,Knaute:2017lll,Klebanov:2007ws}. However, for a mixed state, the EE is not a good measure of correlation and to do so one can study other quantities such as mutual information (MI) which measures the total (classical as well as quantum) correlation between two subsystems $A$ and $B$ which is defined as $I(A,B)=S(A)+S(B)+S(A\cup B)$ where $S$ denotes the entanglement entropy of its associated subsystem \cite{Casini:2004bw,Wolf:2007tdq,Fischler:2012uv,Allais:2011ys,Hayden:2011ag,MohammadiMozaffar:2015wnx,Asadi:2018ijf,Ali-Akbari:2019zkf,DiNunno:2021eyf,Asadi:2018lzr}. When the two subsystems $A$ and $B$ are complementary to each other (total system describes by a pure state), $I(A,B)=2S(A)=2S(B)$.  Since the divergent pieces in the EE cancel out, the MI is a finite scheme-independent quantity and it is always non-negative because of the subadditivity of the EE.

Another important quantity which measure the total correlation between two subsystems $A$ and $B$ in a mixed state is entanglement of purification (EoP) \cite{arXiv:quant-ph/0202044v3,arXiv:quant-ph/1502.01272}. In general, we can purify a mixed state $\rho_{AB}$ to a pure state $\vert \psi_{AA'BB'}\rangle$ by enlarging the Hilbert space. The EoP is defined by the minimum of the EE between two subsystems $AA'$ and $BB'$ among all possible purifications. From this definition it is clear that the EoP between two subsystems $A$ and $B$ reduces to the EE of subsystem $A$ if one considers the subsystem $B$ as
complementary of subsystem $A$. There is a holographic prescription to calculate the EoP in which this quantity is dual to the minimal cross-section of entanglement wedge $Ew$ of $\rho_{AB}$ \cite{Takayanagi:2017knl,Nguyen:2017yqw}. When $\rho_{AB}$ describes a pure state, $Ew$ between subsystems $A$ and $B$ reduces to the holographic entanglement entropy (HEE) of subsystem $A$. There are other correlation measures such as reflected
entropy, odd entropy and logarithmic negativity which are discussed in the literature and holographically all of them related to $Ew$ \cite{Dutta:2019gen,Tamaoka:2018ned,Kudler-Flam:2018qjo,Basu:2021awn,Basak:2022cjs,Basu:2022nds,Vasli:2022kfu}. Various aspects of the EoP are discussed in the literature \cite{Bhattacharyya:2018sbw,Caputa:2018xuf,Camargo:2020yfv,Yang:2018gfq,Espindola:2018ozt,Bao:2017nhh,Umemoto:2018jpc,Bao:2018gck,Bao:2018fso,
Bhattacharyya:2019tsi,Liu:2019qje,Jokela:2019ebz,BabaeiVelni:2019pkw,BabaeiVelni:2020wfl,Amrahi:2020jqg,Amrahi:2021lgh,Sahraei:2021wqn,Guo:2019azy,
Ghodrati:2019hnn,Guo:2019pfl,Bao:2019wcf,Harper:2019lff,Camargo:2021aiq,Saha:2021kwq,Chowdhury:2021idy,ChowdhuryRoy:2022dgo,Jain:2020rbb}.

The remainder of this paper is organized as follows. In section \ref{back}, we review the background and describe the thermodynamics of the considered model.  In section \ref{measure}, after a short review on HEE, we introduce the MI and EoP and their holographic duals. In section \ref{numerical} we explain our numerical results and describe how the HMI and EoP characterize various phase structures of the system. Finally, we will conclude in section \ref{conclution} with the discussion of our results.
\section{The background and the thermodynamics of the system}\label{back}
As we mentioned in the introduction, we are interested in studying the phase structure of the (3+1)-dimensional non-conformal field theory using the HMI and EoP.  Therefore, we present a review on the five dimensional gravitational theory which is dual to the aforementioned field theory \cite{Gubser:2008ny,Gubser:2008yx} and study the thermodynamics of this model.
\subsection{The review of the background}
We consider black hole solutions which follow from the following 5-dimensional Einstein-dilaton action
\begin{align}\label{action}
\mathcal{S}=\frac{1}{16\pi G_N^{(5)}}\int d^5x\sqrt{-g}\left[\mathcal{R}-\frac{1}{2}(\partial\phi)^2-V(\phi)\right],
\end{align}
where $G_N^{(5)}$ is the $5$-dimensional Newton constant and we consider $16\pi G_N^{(5)}=1$ in the following discussion. $g$ and $\mathcal{R}$ are the determinant of the metric and its corresponding Ricci scalar, respectively. $\phi$ denotes dilaton field and $V(\phi)$ is the dilaton potential which is considered in the following form \cite{Janik:2015waa,Janik:2016btb}
\begin{align}
V(\phi)=-12 \cosh(\gamma \phi)+b_2\phi ^2+b_4 \phi ^4 +b_6 \phi ^6.
\end{align}
Each set of parameters $\lbrace \gamma , b_2 , b_4 , b_6 \rbrace$ leads to a black hole solution. This family of black hole solutions has been introduced in \cite{Gubser:2008ny,Gubser:2008yx} and used to mimic the equation of state of QCD. For small $\phi$, one can expand $V(\phi)$ as follows
\begin{align}
V(\phi)\sim -12+ \frac{1}{2}m^2\phi ^2 +\mathcal{O}(\phi ^4),
\end{align}
where  $m^2\equiv 2(b_2-6\gamma ^2)$. The first term is the negative cosmological constant (we fix the AdS radius to one). According to the AdS/CFT dictionary, the scalar field $\phi$ in the bulk is dual to a scalar operator in the dual boundary field theory $O_\phi$ which is known as field-operator correspondence  \cite{Witten:1998qj,CasalderreySolana:2011us}. The conformal dimension of the scalar operator, $\Delta$, is related to the mass of the scalar field in the bulk through $\Delta (\Delta -4)=m^2$. Holographically, the presence of the dilaton field in the bulk is dual to a deformation of  the boundary conformal field theory 
\begin{align}
\mathcal{L}= \mathcal{L}_{\text{CFT}}+\Lambda^{4-\Delta}O_{\phi},
\end{align}
where $\Lambda$ is an energy scale and $\Delta$ is considered to be at range $2\leqslant \Delta <4$ corresponding to the relevant deformation of the CFT and satisfies the Breitenlohner-Freedman bound $m^2\geqslant -4$ \cite{Breitenlohner:1982bm,Breitenlohner:1982jf}. 

\begin{table}[]
\centering
\captionsetup{singlelinecheck=false,font={small}}
\caption{The three scalar potential $V_{\rm QCD}$,  $V_{2{\rm nd}}$ and  $V_{1{\rm st}}$}
\label{tab1-1}
\begin{tabular}{rrrrrrr}
    \hline
         potential & \ \ \ \ \ $\gamma$ & \ \ \ \  $b_2$ & \ \ \ \   $b_4$ & \ \ \ \ $b_6$ & \ \ \ \ $\Delta$\ \\
    \hline
         $V_{\rm QCD}$ & \ \ \ \ \ \ \ 0.606 & \ \ \ \ 1.4 & \ \ \ $-0.1$ & \ \ \ \ 0.0034 & \ \ \ \ 3.55 \\

         $V_{2{\rm nd}}$ \ \ & \ \ \ \ \ \ \ $\frac{1}{\sqrt{2}}$ & \ \ \ \ 1.958 & \ \ \ \ 0 & \ \ \ \ 0 & \ \ \ \ 3.38 \\

         $V_{1{\rm st}}$ \ \ & \ \ \ \ \ \ \  $\sqrt{\frac{7}{12}}$ & \ \ \ \  2.5 & \ \ \ \ 0 & \ \ \ \ 0 & \ \ \ \ 3.41 \\
    \hline 
\end{tabular} 
\end{table}

Different choices of the parameters $\lbrace \gamma , b_2 , b_4 , b_6 \rbrace$  lead to different thermodynamical properties of this model. We consider three sets of parameters for dilaton potential, labeled by $V_{\rm QCD}$, $V_{2{\rm nd}}$ and $V_{1{\rm st}}$ which are given in Table \ref{tab1-1}. The parameters for the $V_{\rm QCD}$  are chosen to fit the lattice QCD (lQCD) data from  \cite{Borsanyi:2012cr} and from the QCD phase diagram we know that at zero chemical potential, the system exhibits a crossover phase transition at a critical temperature. The parameters for the  $V_{1{\rm st}}$ and $V_{2{\rm nd}}$ are chosen so that the system exhibits the first and the second order phase transitions at a certain critical temperature, respectively.

The equations of motion following from the action \eqref{action} are 
\begin{subequations}\label{EoM1}\begin{align}
R_{\mu\nu}-\frac{1}{2}\nabla_\mu \phi \nabla_\nu \phi-\frac{1}{3}V(\phi)g_{\mu\nu}=0, \\ \nabla_\mu\nabla^\mu\phi-\frac{dV(\phi)}{d\phi}=0.
\end{align}\end{subequations}
By considering the following ansatz in the $r=\phi$ gauge \cite{Gubser:2008ny,Gubser:2008yx}
\begin{align}\label{metric}
ds^2=e^{2A(\phi)}\left(-h(\phi) dt^2+d\vec{x}^2\right)+e^{2B(\phi)}\frac{d\phi ^2}{h(\phi)},
\end{align}
the equations of motion \eqref{EoM1}  become simple and are given by
\begin{subequations}\label{EoM2}\begin{align}
A''-A'B'+\frac{1}{6}=0 ,\\
h''+(4A'-B')h'=0 , \\
6A'h'+h(24A'^2-1)+2e^{2B}V=0 , \\
4A'-B'+\frac{h'}{h}-\frac{e^{2B}}{h}V'=0,
\end{align}\end{subequations}
where $X'=\frac{dX}{d\phi}$. The strongly coupled field theory lives on the boundary at $\phi\rightarrow 0$ and the location of the black hole horizon is determined by the simple zero of the blackening function $h(\phi)$
\begin{align}\label{horizon}
h(\phi_H)=0.
\end{align}
In order to solve \eqref{EoM2} we review the method proposed in \cite{Gubser:2008ny,Gubser:2008yx} known as master equation formalism. If we consider $ G(\phi)\equiv A'(\phi)$ the solution of the equations of motion can be expressed as
\begin{subequations}\label{EoM3}\begin{align}
A(\phi)&=A_H+\int_{\phi_H}^\phi d\tilde{\phi}G(\tilde{\phi}),\\
B(\phi)&=B_H+\ln\left(\frac{G(\phi)}{G(\phi_H)}\right)+\int_{\phi_H}^\phi \frac{d\tilde{\phi}}{6G(\tilde{\phi})},\\
h(\phi)&=h_H+h_1\int_{\phi_H}^\phi d\tilde{\phi}e^{-4A(\tilde{\phi})+B(\tilde{\phi})},
\end{align}\end{subequations}
where $A_H$, $B_H$, $h_H$ and $h_1$ are the integration constants which are obtained by applying appropriate boundary conditions at the black hole horizon \eqref{horizon}  and the boundary of this background $\phi \rightarrow 0$ \cite{Gubser:2008ny,Gubser:2008yx}
\begin{subequations}\label{EoM4}
\begin{align}
A_H&=\frac{\ln (\phi_H)}{\Delta -4}+\int_0^{\phi_H}d\phi\left[G(\phi)-\frac{1}{(\Delta -4)\phi}\right],\\
B_H&=\ln\left(-\frac{4V(\phi_H)}{V(0)V'(\phi_H)}\right)+\int_0^{\phi_H} \frac{d\phi}{6G(\phi)},\\
h_H&=0,\\
h_1&=\frac{1}{\int_{\phi_H}^0 d\phi e^{-4A(\phi)+B(\phi)}}.
\end{align}
\end{subequations}
By manipulating the field equations \eqref{EoM2} one can derive the following ``nonlinear master equation"
\begin{align}\label{master}
\frac{G'}{G+V/3V'}=\frac{d}{d\phi}\ln\left(\frac{G'}{G}+\frac{1}{6G}-4G-\frac{G'}{G+V/3V'}\right).
\end{align}
By solving this equation and obtaining $G(\phi)$, the metric coefficients $A(\phi)$, $B(\phi)$ and $h(\phi)$ can be obtained from \eqref{EoM3}. In order to solve \eqref{master} we need appropriate boundary conditions. From \eqref{master} one can find the following near-horizon expansion
\begin{align}\label{cond1}
G(\phi)=-\frac{V(\phi_H)}{3V'(\phi_H)}+\frac{1}{6}\left(\frac{V(\phi_H)V''(\phi_H)}{V'(\phi_H)^2}-1\right)(\phi-\phi_H)+\mathcal{O}(\phi-\phi_H)^2.
\end{align}
This expansion implies that
\begin{align}\label{cond2}
G'(\phi_H)=\frac{1}{6}\left(\frac{V(\phi_H)V''(\phi_H)}{V'(\phi_H)^2}-1\right).
\end{align}
Unfortunately, it is hard to solve \eqref{master} analytically and hence we use numerical methods to solve it. Our strategy is to specify a value for $\phi_H$ and use the boundary conditions \eqref{cond1} and \eqref{cond2} for integrating \eqref{master}. Each value of $\phi_H$ leads to a unique black hole solution. We obtain numerically a family of black hole solutions for different values of $\phi_H$.
\subsection{The thermodynamics}\label{thermo}
In this subsection we would like to study the thermodynamics of the field theory which is dual with the black hole background described by metric \eqref{metric}. The Hawking temperature can be obtained from the metric \eqref{metric} 
\begin{align}\label{temp}
T=\frac{e^{A(\phi_H)-B(\phi_H)}|h'(\phi_H)|}{4\pi},
\end{align}
and the Beckenstein-Hawking formula for entropy density leads to 
\begin{align}\label{entropy}
s=\frac{e^{3A(\phi_H)}}{4}.
\end{align}
At zero chemical potential the speed of sound and the specific heat are related to the temperature and the entropy density as 
\begin{align}\label{sp}
c_s^2=\frac{d\ln T}{d\ln s}=\frac{d\ln T/d\phi_H}{d\ln s/d\phi_H}, \ \ \ \ \ \ \ \ \ \ \ \ \ \ \ \ C_v=T\frac{\partial s}{\partial T}.
\end{align}
One can obtain the free energy from the first law of thermodynamics, $dF=-sdT$. At zero chemical potential the free energy of a system with constant volume are given by
\begin{align}
F(T)=F(T_0)-\int_{T_0}^Ts(\tilde{T})d\tilde{T}.
\end{align}
We take $T_0$ to be zero. Since the metric of the zero temperature background (thermal gas) coincides with the black hole metric in the limit $\phi_H\rightarrow \infty$, we expect that the free energy of the black hole background vanishes in this limit, $F(T_0=0)=F(\phi_H\rightarrow \infty)=0$ and hence the free energy is given by
\begin{align}\label{free}
F(\phi_H)=-\frac{1}{4}\int_{\phi_H}^{\infty} e^{3A(\tilde{\phi}_H)}\frac{dT(\tilde{\phi}_H)}{d\tilde{\phi}_H} d\tilde{\phi}_H.
\end{align}
\begin{figure}[]
\centering
\subfloat[$V_{\rm QCD}$]{\includegraphics[scale=0.33]{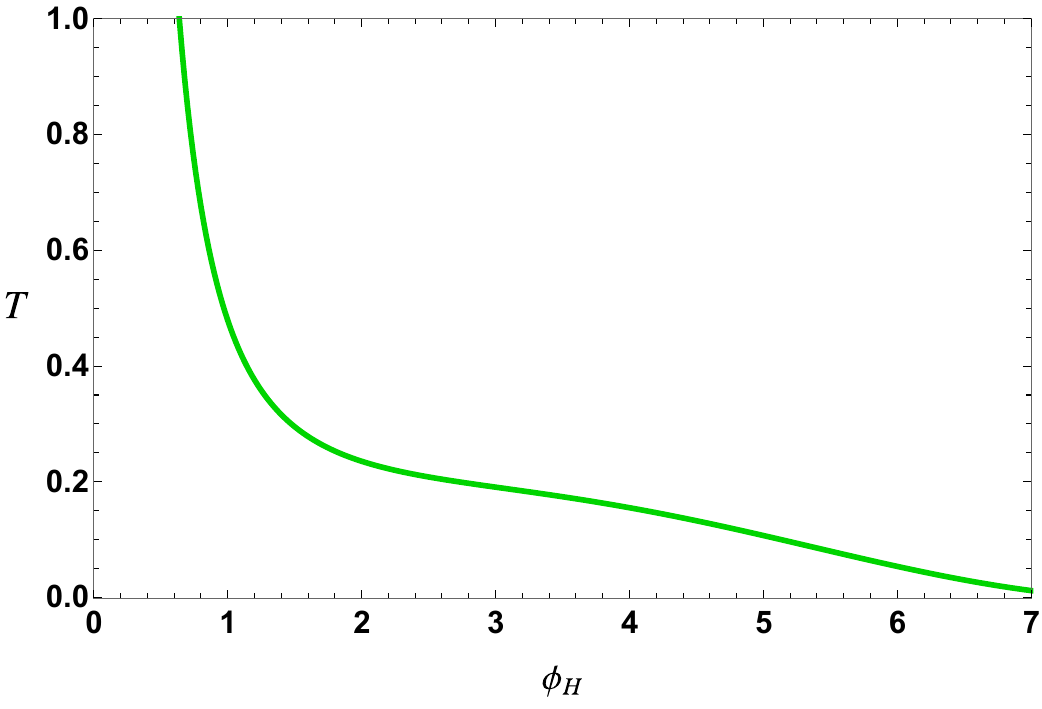} }
\subfloat[$V_{2{\rm nd}}$]{\includegraphics[scale=0.33]{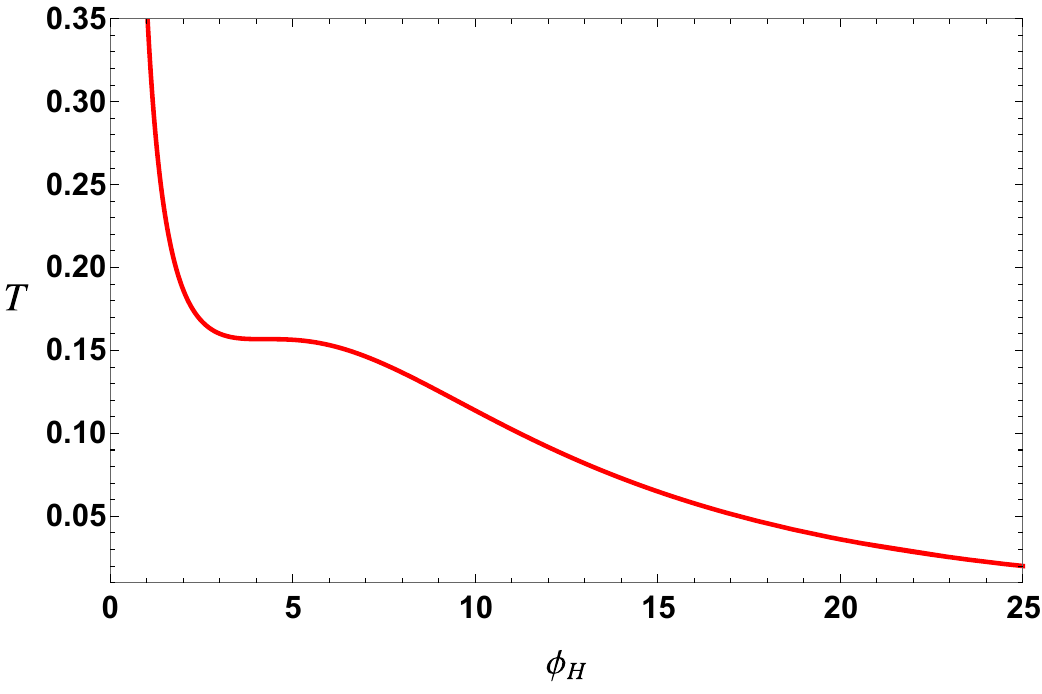} }
\subfloat[$V_{1{\rm st}}$]{\includegraphics[scale=0.33]{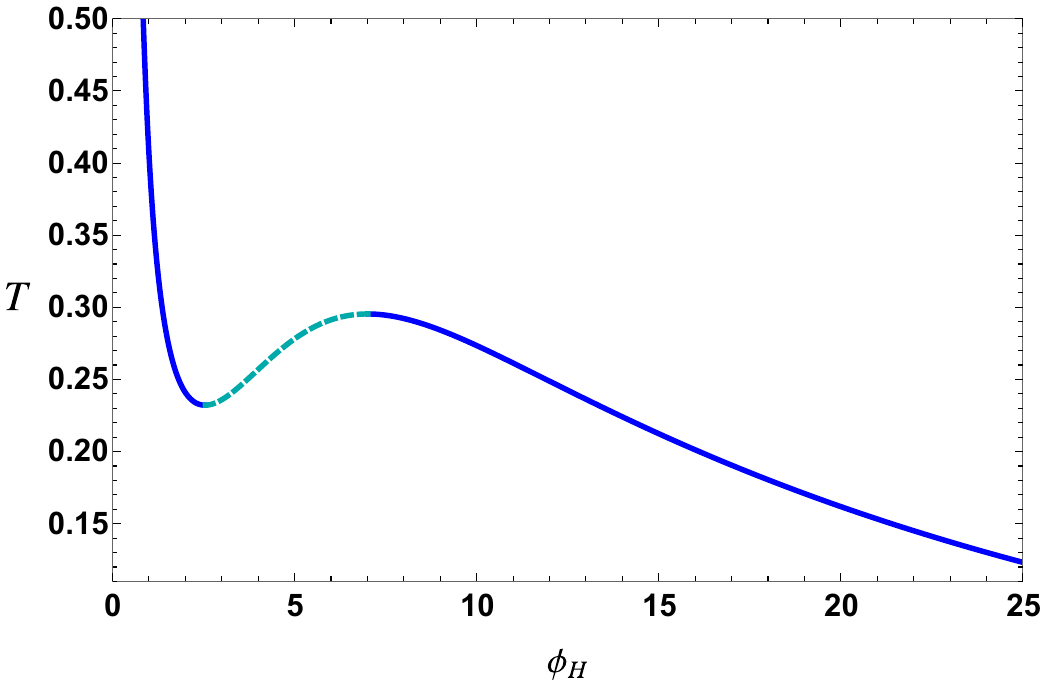}}
\captionsetup{justification=raggedright,singlelinecheck=false,font={small}}
\caption{The temperature $T$ v.s. horizon $\phi_H$ for the dilaton potentials $V_{\rm QCD}$ (left), $V_{2{\rm nd}}$ (middle) and $V_{1{\rm st}}$ (right).}\label{T-over-phiH}
\label{}
\end{figure}
We plot the temperature $T$ v.s. horizon $\phi_H$ in FIG. \ref{T-over-phiH}. From the left diagram, related to $V_{\rm QCD}$, we observe that the temperature monotonically decreases to zero. In this case the black hole solutions are always thermodynamically stable\footnote{If the thermodynamic potential of a system is given by $\psi(x_1,...,x_n)$ which depends on some set of variables $\lbrace x_1,...,x_n \rbrace$, the Hessian matrix $H$ of the associated potential defined by $H_{ij}\equiv \left[\frac{\partial^2 \psi}{\partial x_i \partial x_j}\right]$. The system is called stable when the Hessian matrix is positive-definite.} and a smooth crossover phase transition happens there.  In the middle diagram where we consider $V_{2{\rm nd}}$, there is a critical point at $(\phi_c^{2nd}\simeq 4.27, T_c\simeq 0.1568)$ which $\frac{dT}{d\phi_H}$ becomes zero at this point  and it is a point of inflection. In subsubsection \ref{2nd-thermo} we will discuss that the system possesses a second order phase transition at this point. In the Right diagram, corresponding to $V_{1{\rm st}}$, it is shown that the temperature has a local minimum $T_{min}\equiv T_m\simeq 0.2321$ and a local maximum $T_{max}\simeq0.2954\simeq 1.272T_m$ at $\phi_H=\phi_{min}^{1{\rm st}}\simeq2.54$ and $\phi_H=\phi_{max}^{1{\rm st}}\simeq7.01$, respectively. Between $\phi_{min}^{1{\rm st}}$ and $\phi_{max}^{1{\rm st}}$ the black hole solutions are thermodynamically unstable. We expect a Hawking-Page phase transition happens at a temperature between $T_{min}$ and $T_{max}$. We will discuss it later in subsubsection \ref{1st-thermo}.

In the following, we will discuss the details of thermodynamics properties of the dual field theory for dilaton potentials $V_{\rm QCD}$, $V_{2{\rm nd}}$ and $V_{1{\rm st}}$, respectively.
\subsubsection{$V_{\rm QCD}$}
\begin{figure}[]
\centering
\subfloat[]{\includegraphics[scale=0.35]{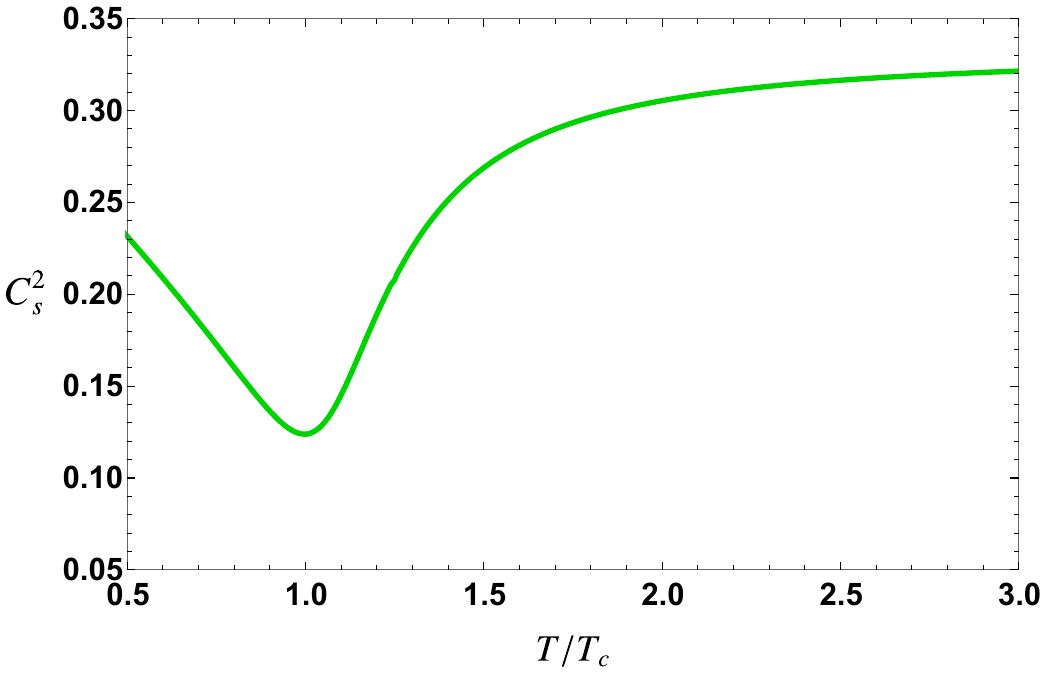}\label{a1}} \ \ \ \ \
\subfloat[]{\includegraphics[scale=0.35]{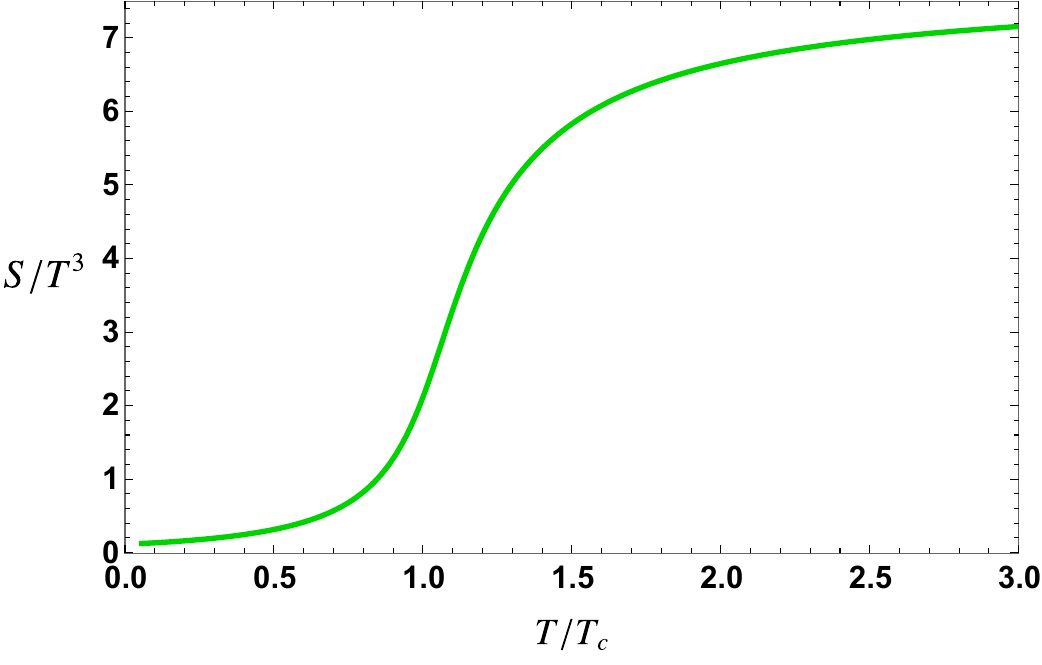}\label{a2}} \\
\subfloat[]{\includegraphics[scale=0.36]{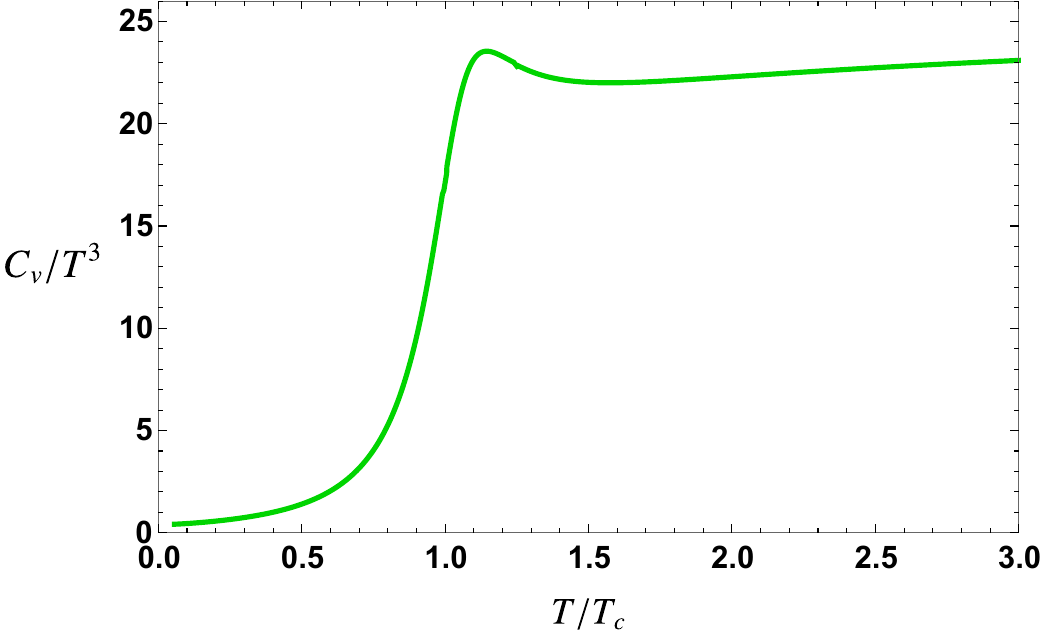} \label{a3}} \ \
\subfloat[]{\includegraphics[scale=0.36]{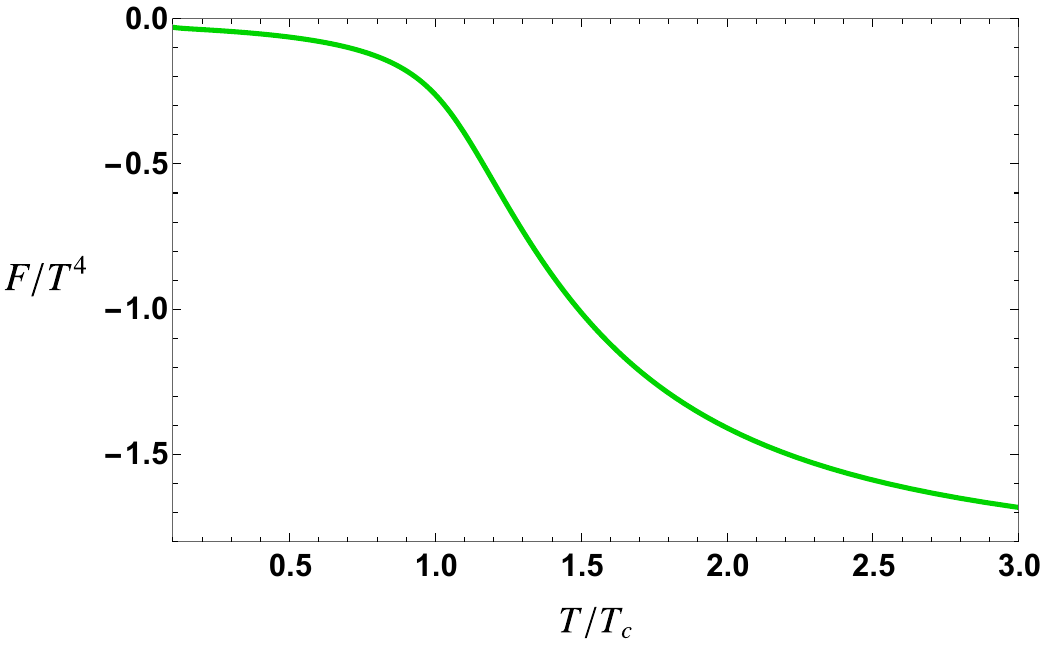}\label{a4}} \ \ \
\captionsetup{justification=raggedright,singlelinecheck=false,font={small}}
\caption{The square of speed of sound $c_s^2$ (a), entropy density $s$ (b), specific heat $C_v$ (c) and free energy $F$ (d) v.s. temperature $T$ for $V_{\rm QCD}$ potential.}\label{CO-thermo}
\label{}
\end{figure}

In FIG. \ref{CO-thermo} we show the dependence of the entropy density $s$, the square of speed of sound $c_s^2$, specific heat $C_v$ and free energy $F$ in terms of temperature $T$ for $V_{\rm QCD}$. The dependence of  $c_s^2$ on $T$ is in complete agreement with the lattice QCD results \cite{Borsanyi:2012cr}. There is a critical temperature $T_c$ corresponding to the lowest dip of the $c_s^2$ and it is clearly seen that $s$ and $\frac{ds}{dt}\propto C_v$ are both continuous at temperature $\frac{T}{T_c}=1$.
%According to \cite{Finazzo:2014cna} this value should be $143.8$ MeV for QCD which in the unit of this paper this temperature equals to $T_b\simeq 0.181$. From the entropy density v.s. temperature diagram one can see that $S/T^3$ tends to a constant value in the limit of $T\rightarrow \infty$ which corresponds to the conformal field theory. Since in the conformal field theory $S/T^3$ counts the number of degrees of freedom (DOF) we can say that $S/T^3$ counts the effective number of DOF in the non-conformal field theory.
 The free energy of the black hole is always negative and is less than the free energy of the thermal gas. Therefore, the thermodynamic system will always favor the black hole background and the Hawking-Page phase transition does not occur which means that the system has a crossover phase transition. From FIG. \ref{a1} it is seen that in the limit $T\rightarrow\infty$, the value of $c_s^2$ approaches to its conformal value $c_s^2=\frac{1}{3}$ which is expected. This result is also valid for the potentials $V_{2{\rm nd}}$ and $V_{1{\rm st}}$.
\subsubsection{$V_{2{\rm nd}}$}\label{2nd-thermo}
\begin{figure}[]
\centering
\subfloat[]{\includegraphics[scale=0.35]{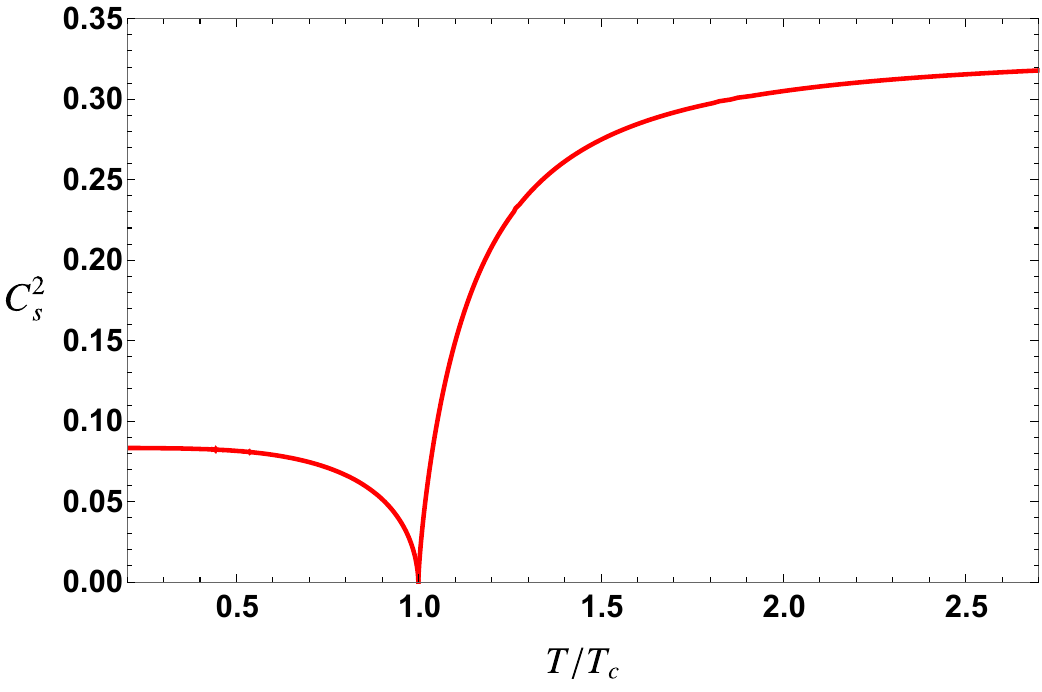}\label{b1}} \ \ \ \ \
\subfloat[]{\includegraphics[scale=0.35]{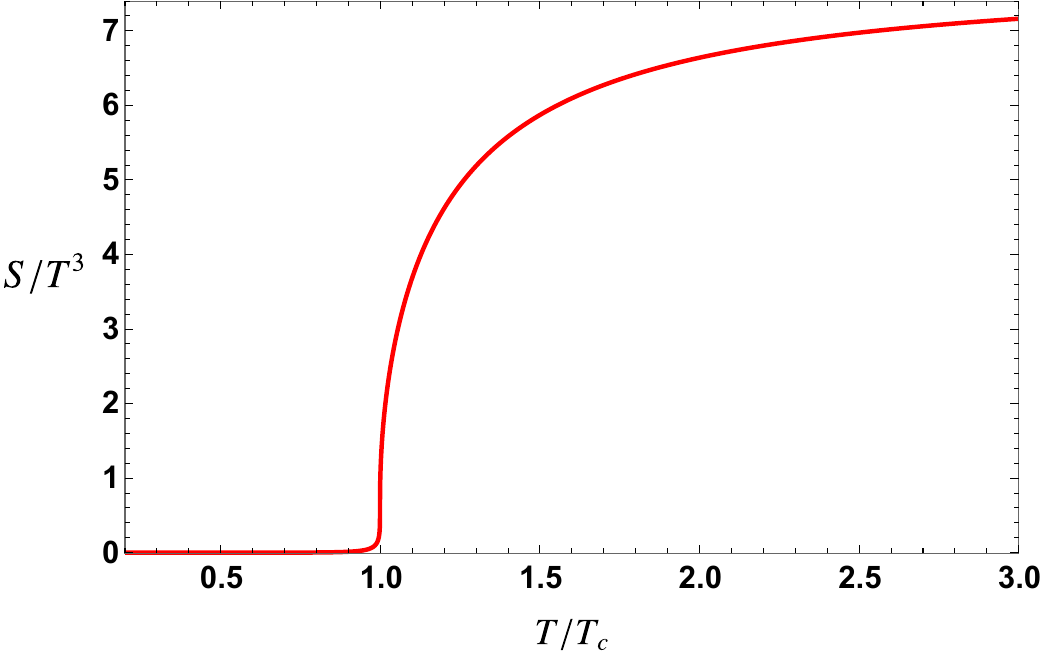}\label{b2}} \\
\subfloat[]{\includegraphics[scale=0.36]{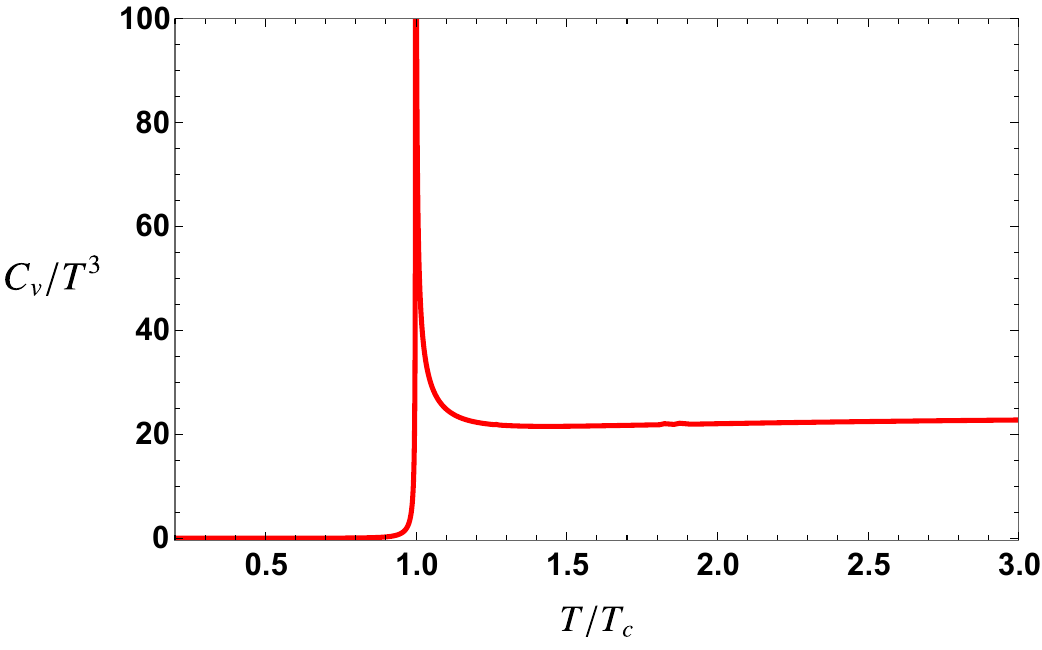} \label{b3}}  \ \
\subfloat[]{\includegraphics[scale=0.36]{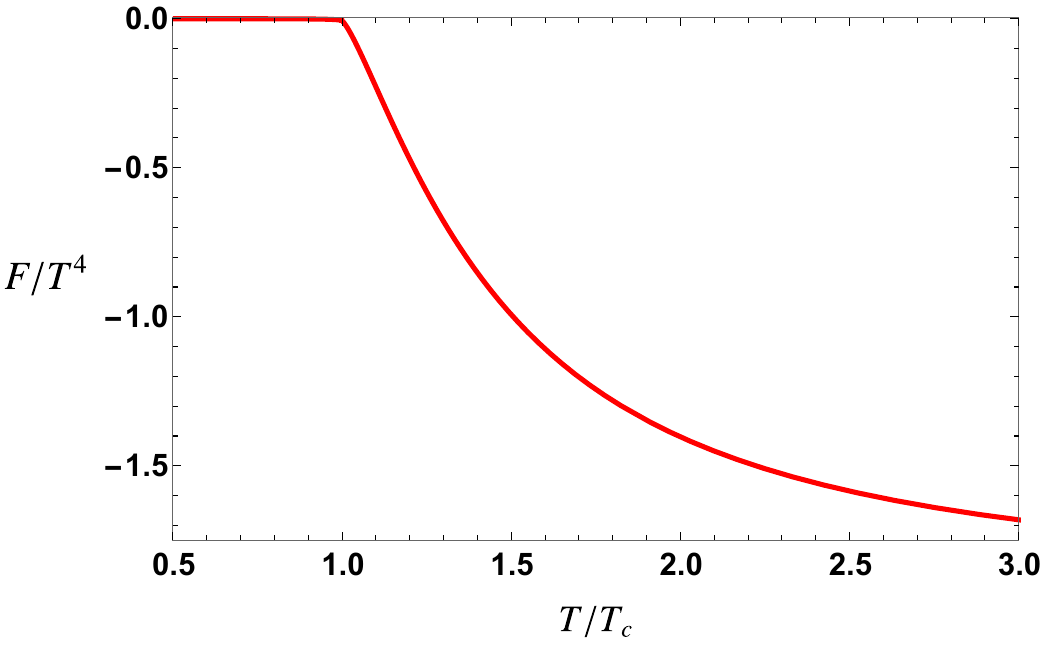}\label{b4}} \ \ \ 
\captionsetup{justification=raggedright,singlelinecheck=false,font={small}}
\caption{The square of speed of sound $c_s^2$ (a), entropy density $s$ (b), specific heat $C_v$ (c) and free energy $F$ (d) v.s. temperature $T$ for $V_{2\rm nd}$ potential.}\label{2nd-thermod}
\end{figure}

\begin{figure}
\centering
\includegraphics[scale=0.4]{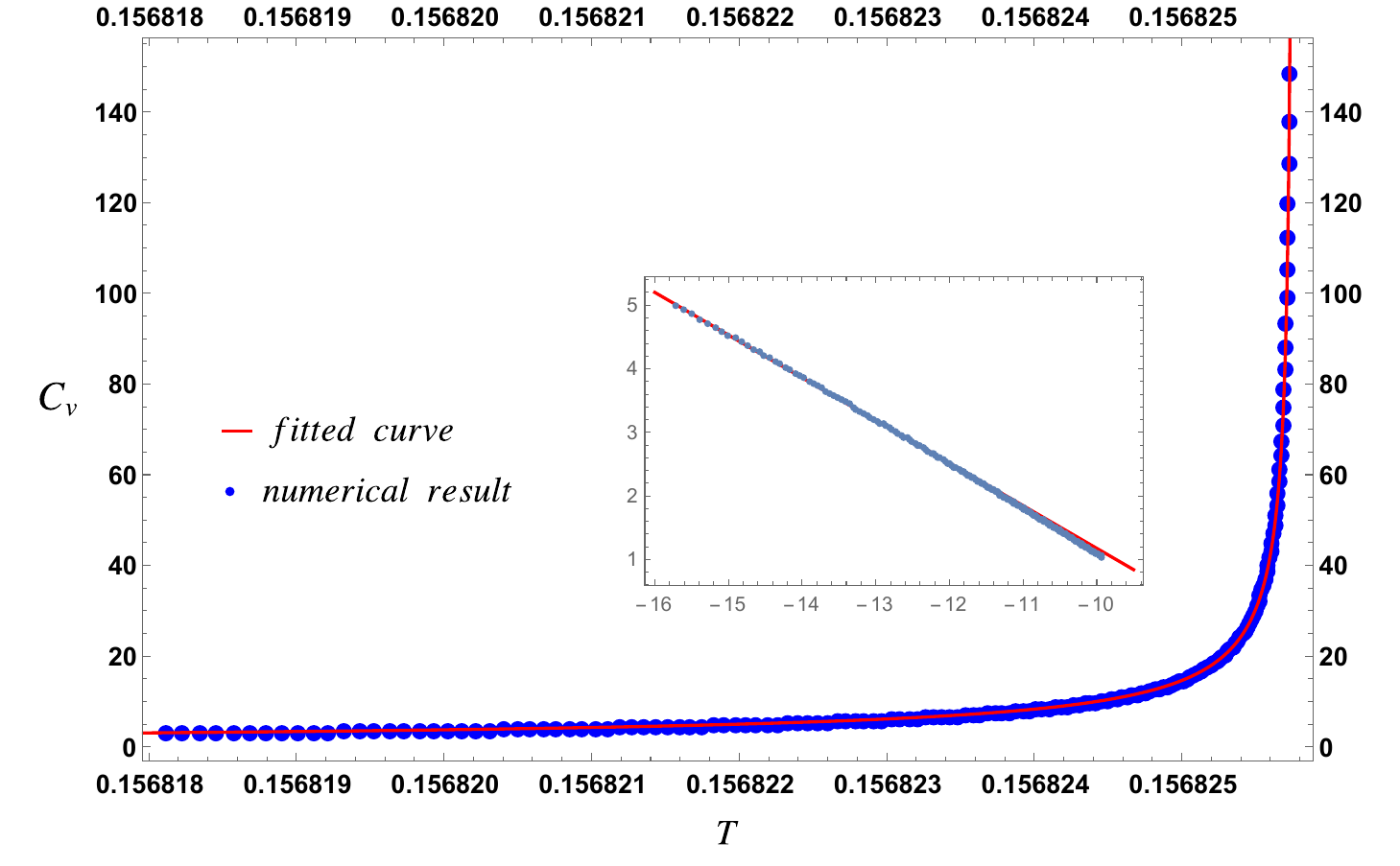} 
\captionsetup{justification=raggedright,singlelinecheck=false,font={small}}
\caption{The specific heat $C_v$ with respect to temperature $T$. The fitted curve with the numerical result is $C_v=0.00396 \ t^{-0.67}$. The corresponding RE and RMS are $0.014$ and $0.19$ respectively. The small plot is the logarithm of the data results and the linear function fitted with them.}
\label{specific-heat-exponent}
\end{figure}

In FIG. \ref{2nd-thermod} the dependence of $s$, $c_s^2$, $C_v$ and $F$ on $T$ for $V_{2{\rm nd}}$ are depicted. $c_s^2$ v.s. $T$ shows that at a critical temperature $T_c$, $c_s^2$ goes to zero but never becomes negative. $F$ and $\frac{dF}{dT}=s$ are finite and continuous at the critical temperature $T_c$ but the value of $\frac{d^2F}{dT^2}\propto C_v$ diverges at $T_c$ and hence the phase transition is second order. The entropy density drops quickly as the temperature approaches to $T_c$. Near the critical temperature, $s$ and the $C_v$ of the system take typically the form
\begin{align}
S(T)\simeq S_0+S_1 t^{1-\alpha}, \ \ \ \ \ \ \ \ \ \ \ C_v(T)\sim t^{-\alpha},
\end{align}
where $t\equiv\frac{\vert T-T_c\vert}{T_c}$ and $\alpha$ is the specific heat critical exponent. In order to find the critical exponent, we focus on the region near the critical point and plot $C_v(T)$ in this region in FIG. \ref{specific-heat-exponent}. By fitting a curve with the numerical result, the value of the critical exponent is found to be $\alpha=0.67$. This value is in complete agreement with the one reported in \cite{Gubser:2008ny,Janik:2016btb}. To get this number, we also plot the linear log-log diagram for which the critical exponent is the slope of a line i.e. $\log (C_v)\propto \alpha \log (t)$. To report how well our result is, we calculate relative error (RE) and root mean square (RMS) which are defined as
\begin{align}
&{\rm RE}=\frac{\alpha-2/3}{2/3}, \cr
&{\rm RMS}=\sqrt{\frac{1}{N}\sum\limits_{i=1}^N\left(y_{\rm fitted}(i)-y_{\rm data}(i)\right)^2},
\end{align}
where $y_{\rm fitted}(i)$ is the value of fitted function $y$ evaluated at $i$th data points $x$, $y_{\rm data}$ is the corresponding value read from data and $N$ is the number of data points. These numbers are reported in the caption of FIG. \ref{specific-heat-exponent}.
\subsubsection{$V_{1{\rm st}}$}\label{1st-thermo}
\begin{figure}[]
\centering
\subfloat[]{\includegraphics[scale=0.35]{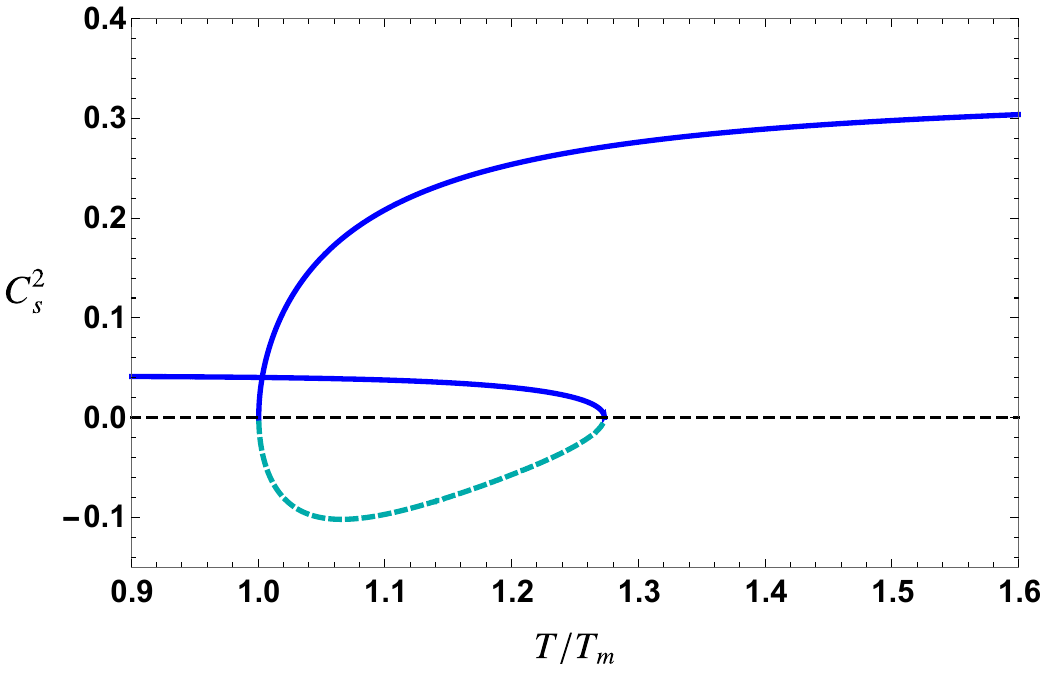}\label{c1}} \ \ \ \ \ 
\subfloat[]{\includegraphics[scale=0.35]{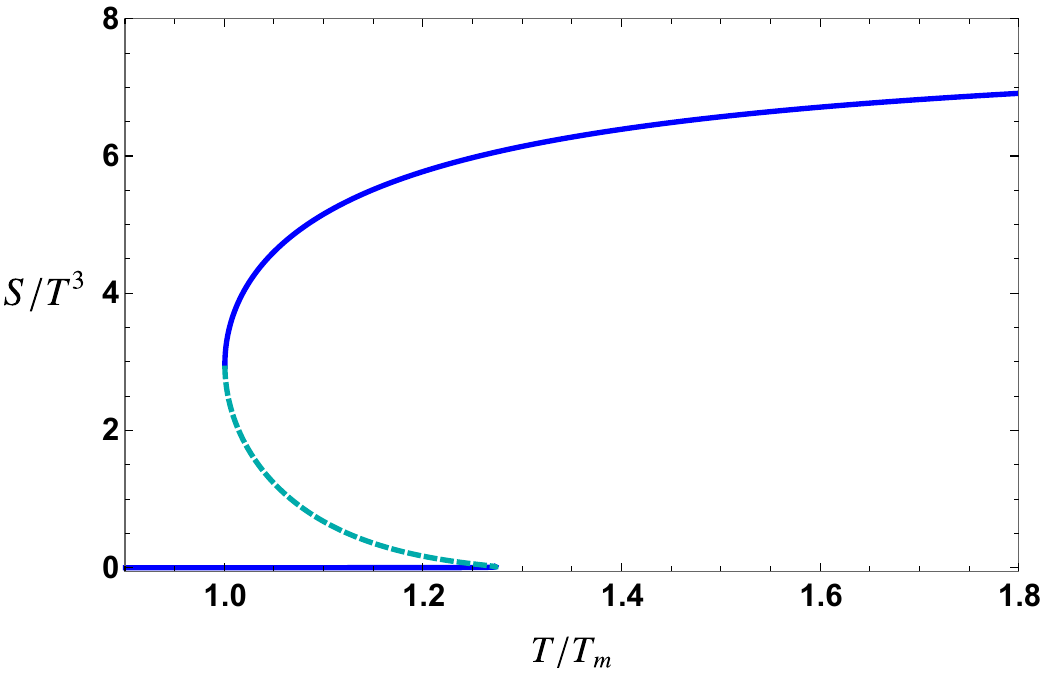}\label{c2}} \\
\subfloat[]{\includegraphics[scale=0.36]{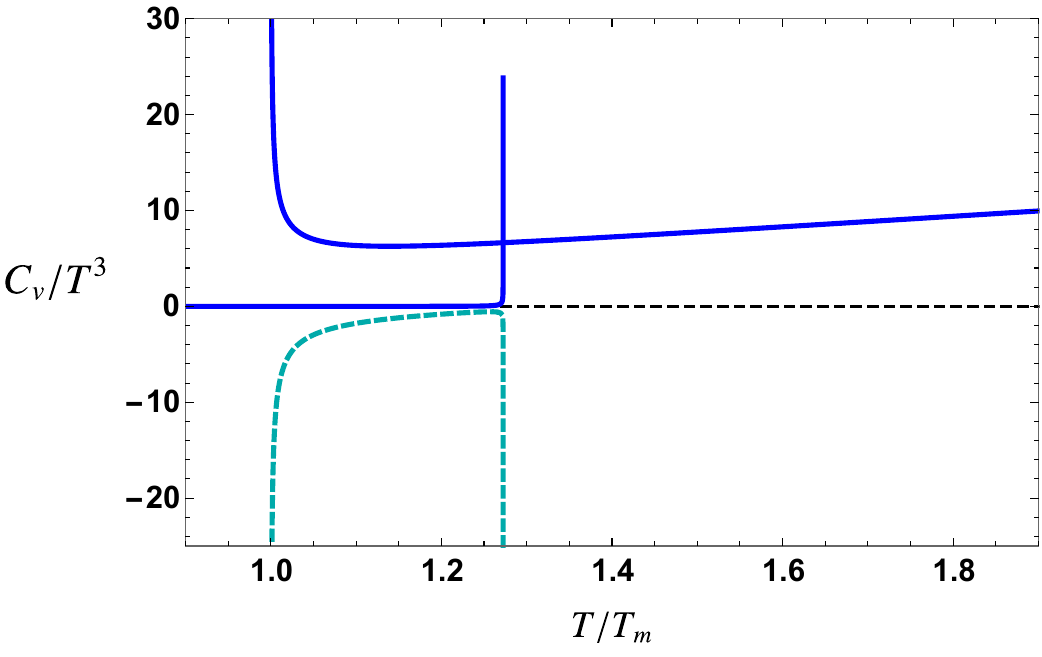} \label{c3}} \ \
\subfloat[]{\includegraphics[scale=0.36]{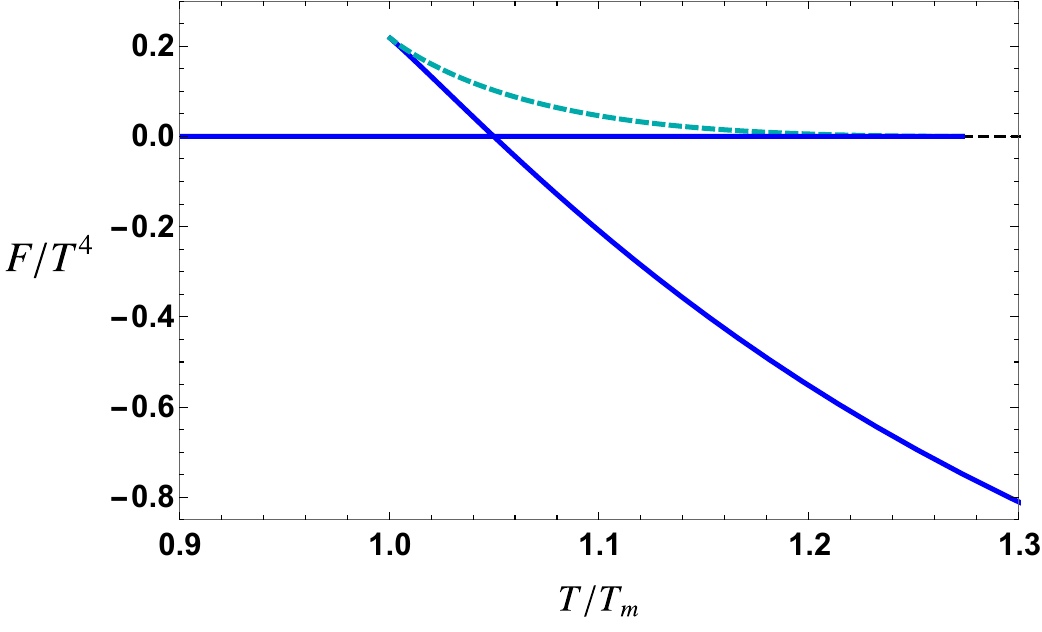}\label{c4}} \ \ \
\captionsetup{justification=raggedright,singlelinecheck=false,font={small}}
\caption{The square of speed of sound $c_s^2$ (a), entropy density $s$ (b), specific heat $C_v$ (c) and free energy $F$ (d) v.s. $T$ for $V_{1\rm st}$ potential. The blue solid curves correspond to the stable black hole solution and the dashed cyan curves correspond to the unstable solution.}\label{1st-thermod}
\end{figure}

In FIG. \ref{1st-thermod}, the dependence of $s$, $c_s^2$, $C_v$ and $F$ on $T$ for $V_{1{\rm st}}$ are shown. From these diagrams one can see that in the range of temperature $T_m< T <1.272T_m$ there are three branches of solutions where two of them correspond to the stable black hole solution (solid blue curves) and one of them corresponds to the unstable black hole solution (dashed cyan curve).  There are no unstable solutions at $T< T_m$ and $1.272T_m< T$. The dashed curves show that the system indicates the Gregory-Laflamme instability \cite{Gregory:1993vy,Gregory:1994bj}  which can be seen from the behavior of $C_v$. In \cite{Gubser:2000ec,Gubser:2000mm,Reall:2001ag} it is shown that in the absence of conserved charges related to gauge symmetries, the Gregory-Laflamme instability is equivalent to the negative values of $C_v$ which is depicted clearly in panel \ref{c3}. From the panels \ref{c1} and \ref{c3} we observe that the negative values of $C_v$ corresponds to the negative $c_s^2$. $F(T)$ shows that the phase transition occurs at critical temperature $T_c\simeq 1.05T_m$ in which the two stable branches of solutions cross each other at this point. Although $F$ is continuous across the phase transition, $\frac{dF}{dT}=s$ is not continuous which means the system possesses a first order phase transition at $T_c$.  Since $s/T^3$ counts the effective number of DOF, this number increases suddenly when we approach to $T_c$ from the left \cite{Zhang:2016rcm}.

\section{The holographic entanglement measures}\label{measure}
In this section we would like to study the HMI and EoP in the black hole background described by metric \eqref{metric} and investigate these entanglement measures behavior near the phase transition temperature. These quantities are known in quantum information theory and measure total, both classical and quantum, correlation between subsystems A and B for the total system described by mixed state $\rho_{AB}$. 
\subsection{The holographic mutual information}
When the system describes by a pure state, the EE is a unique measure which determines the quantum entanglement between subsystem $A$ and its complementary $\bar{A}$. If we consider a pure quantum system described by the density matrix $\rho=\vert \psi \rangle\langle \psi \vert$ and divide the total system into two subsystems $A$ and its complement $\bar{A}$, the entanglement between these subsystems is measured by the EE which is defined as
\begin{align}
S_A=-Tr(\rho_A \log \rho_A),
\end{align}
where $\rho_A=Tr_{\bar{A}}(\rho)$ is  the reduced density matrix for the subsystem $A$. It is shown that the EE contains short-distance divergence which satisfies an area law and hence the EE is a scheme-dependent quantity in the UV limit \cite{Bombelli:1986rw,Srednicki:1993im}. It is difficult to calculate the EE using QFT techniques. However, in the framework of the gauge/gravity duality, there is a simple prescription to compute entanglement entropy in terms of a geometrical quantity in the bulk \cite{Ryu:2006bv,Ryu:2006ef}.  According to this prescription, the holographic entanglement entropy (HEE) is given by
\begin{align}\label{HEE}
S_A=\frac{\rm{Area}(\Gamma_A)}{4G_N^{(d+2)}},
\end{align}
where $\Gamma_A$ is a codimension-2 minimal hypersurface, called Ryu-Takayanagi surface (RT-surface), in the bulk whose boundary coincides with the boundary of region $A$. In \cite{Zhang:2016rcm} the behavior of the HEE is studied in three dilaton potentials, by choosing suitable parameters to the scalar self interaction potential, and they showed that  the HEE can be characterize the crossover/ phase transition. 

If there are two disjoint subsystems on the boundary entangling region, one of the most important quantity to study is the mutual information which measures the total correlation between the two subsystems, including both classical and quantum correlations, in a mixed state \cite{Groisman}. The mutual information between two disjoint subsystems $A$ and $B$ is defined as a linear combination of entanglement entropy
\begin{figure}
\centering
\includegraphics[scale=0.28]{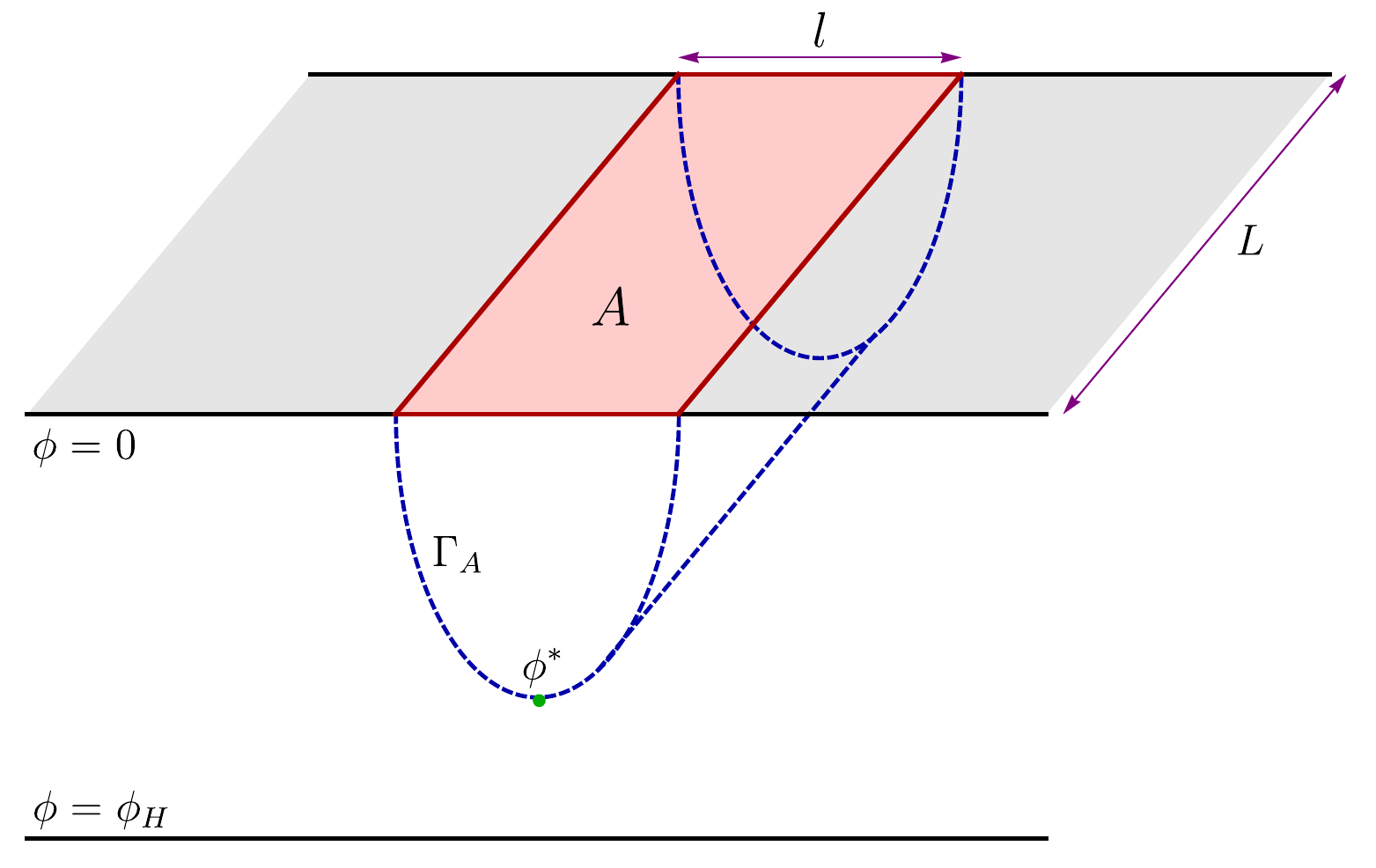}
\includegraphics[scale=0.285]{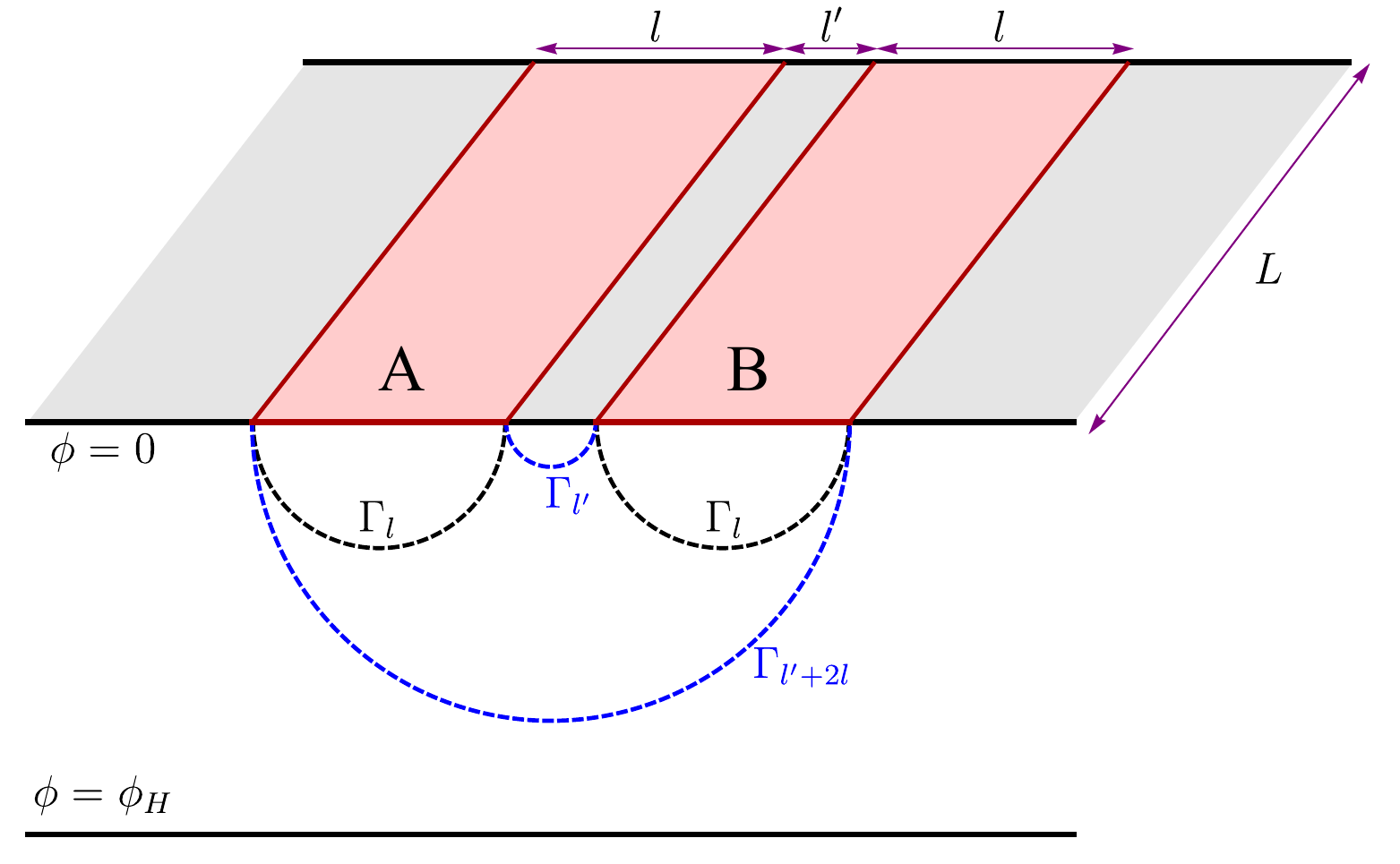}
\captionsetup{justification=raggedright,singlelinecheck=false,font={small}}
\caption{Left: A simplified sketch of a strip region $A$ with width $l$ and length $L$. $\Gamma_A$ is the
RT-surface of the region $A$ and $\phi^*$ is the turning point of this surface.
Right: A simplified sketch of two strip regions $A$ and $B$ with equal size $l$ which are separated by the distance $l'$. When $l'$ is small enough, the minimal surface of $A\cup B$ are given by $\Gamma_{l'}\cup \Gamma_{l'+2l}$ and when  $l'$ is large enough, the minimal surface of $A\cup B$ are given by $2\Gamma_l$. The minimal surfaces, the dashed curves, are denoted by $\Gamma$.}
\label{MIandEE}
\end{figure}
\begin{align}\label{Mut}
I(A,B)=S_A+S_B-S_{A\cup B},
\end{align}
where $S_A$, $S_B$ and $S_{A\cup B}$ denote the entanglement entropy of the region $A$, $B$ and $A\cup B$, respectively. From \eqref{Mut} one can conclude that the mutual information is a finite quantity since the divergent pieces in the entanglement entropy cancel out and it is always positive because of the subadditivity of the entanglement entropy, $S_A+S_B\geq S_{A\cup B}$. We consider the two symmetric disjoint subsystems both rectangular strips of size $l$ which are separated by the distance $l'$ on the boundary, see FIG. \ref{MIandEE}. Using holographic prescription we can compute easily the HEE of the individual subsystems $A$ and $B$. In order to compute $S_{A\cup B}$, we have two possible configurations. When the separation distance is large enough (disconnected configuration), the two subsystems $A$ and $B$ are completely disentangled and we have $S_{A\cup B}=S_A+S_B=2S(l)$, and hence the mutual information vanishes $I(A,B)=0$. On the other hand, when the two subsystems $A$ and $B$ are close enough to each other (connected configuration), $S_{A\cup B}=S(l')+S(l'+2l)$ and we get $I(A,B)>0$. One can assume that the transition of the mutual information from positive values to zero occurs at the distance which we call $x_d$. To summarize, $I(A,B)$ is given by
\begin{align}\label{HMI}
I(A,B)=\Bigg\lbrace
\begin{array}{lr}
2S(l) -S(2l+l')-S(l') \ \ l'<x_d   & \\ 
0 \ \ \ \ \ \ \ \ \ \ \ \ \ \ \ \ \ \ \ \ \ \ \ \ \ \ \ \ \ \ \ \ \ l'\geqslant x_d   & 
\end{array} .
\end{align}

%\begin{figure}[]
%\centering
%\includegraphics[scale=0.33]{1st-MI-over-T} 
%\includegraphics[scale=0.33]{2nd-MI-over-T} 
%\includegraphics[scale=0.33]{CO-MI-over-T}
%\caption{The HMI  v.s. temperature $T$ for the dilaton potentials $V_{\rm QCD}$ (left), $V_{2{\rm nd}}$ (middle) and $V_{1{\rm st}}$ (right).}\label{MI-over-T}
%\label{}
%\end{figure}
\subsection{The holographic entanglement of purification}
When the system is described by a mixed state, another important quantity to study is the entanglement of purification which measures the total (quantum and classical) correlation between two disjoint subsystems. In order to define the EoP, we consider a mixed bipartite system with density matrix $\rho_{AB}$. we can always purify this mixed state into a pure state $\vert \psi_{AA'BB'}\rangle$ by adding auxiliary degrees of freedom to the Hilbert space as $\mathcal{H}_A\otimes \mathcal{H}_B\rightarrow \mathcal{H}_A\otimes \mathcal{H}_B \otimes \mathcal{H}_{A'}\otimes \mathcal{H}_{B'}$ such that the total density matrix in enlarged Hilbert space is given by $\rho_{AA'BB'}=\vert \psi_{AA'BB'}\rangle\langle\psi_{AA'BB'}\vert$. This pure state is called a purification of  $\rho_{AB}$ if we have
$\rho_{AB}=Tr_{A'B'}\left(\vert \psi_{AA'BB'}\rangle\langle\psi_{AA'BB'}\vert\right)$. The EoP is then defined by minimizing the entanglement entropy $S_{AA'}$ over all purifications of $\rho_{AB}$ \cite{arXiv:quant-ph/0202044v3}
\begin{align}
E_p(\rho_{AB})=\underset{\vert \psi _{AA'BB'}\rangle}{\rm{min}}(S_{AA'}),
\end{align}
where $S_{AA'}$ is the entanglement entropy corresponding to the density matrix $\rho_{AA'}={\rm{Tr}}_{BB'}\big[\left(|\psi\rangle_{ABA'B'}\right) \left({}_{ABA'B'}\langle\psi|\right)\big]$.
\begin{figure}
\centering
\includegraphics[scale=0.45]{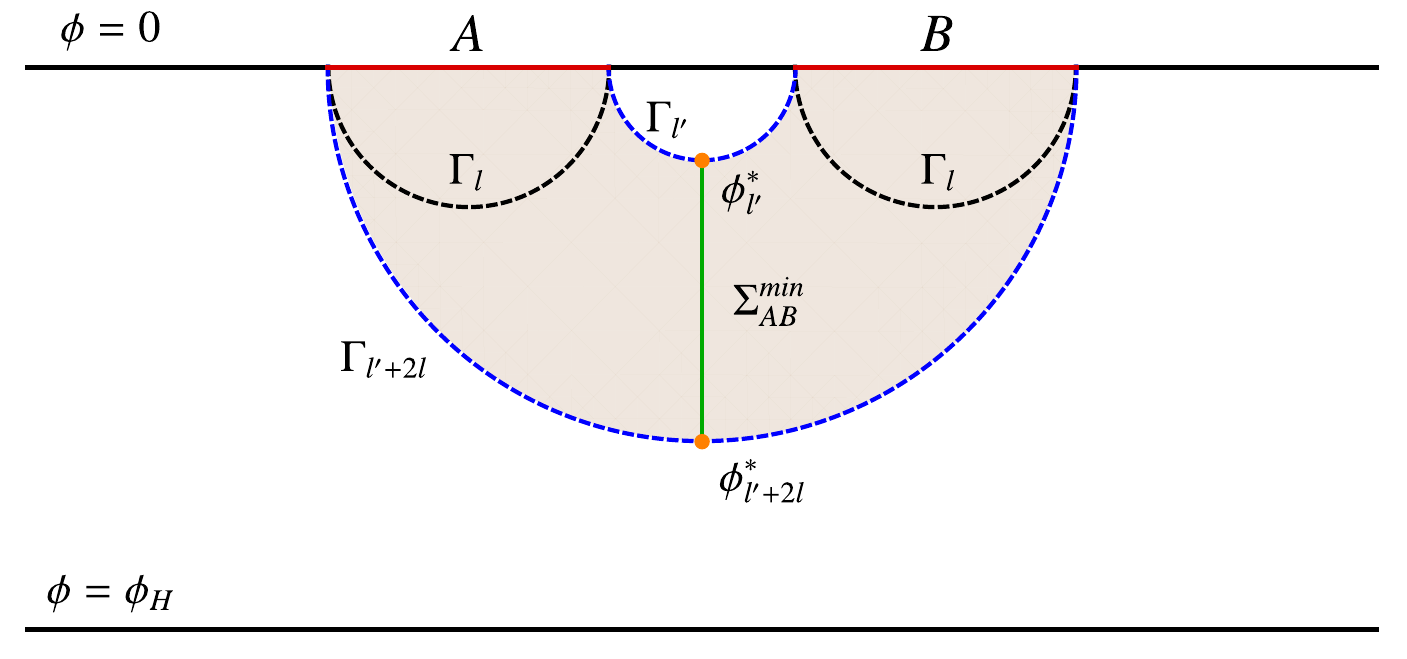}
\captionsetup{justification=raggedright,singlelinecheck=false,font={small}}
\caption{The gray region shows the entanglement wedge dual to $\rho_{AB}$ and $\Sigma _ {AB}^{min}$, the green curve, is the entanglement wedge cross section between subregions A and B. Here we only show the connected configuration. For disconnected configuration the entanglement wedge cross section vanishes $E_w=0$. The RT-surfaces, the dashed curves, are denoted by $\Gamma$.}
\label{EWCS1}
\end{figure}
In general, it is a difficult task to compute the EoP in the context of the QFT. Holographically, it has been conjectured  that the EoP is dual to the entanglement wedge cross-section $E_w$ of $\rho_{AB}$ which is defined by \cite{Takayanagi:2017knl, Nguyen:2017yqw}
\begin{align}\label{EWCS}
E_w(\rho_{AB})=\frac{{\rm{Area}}(\Sigma_{AB}^{min})}{4G_N^{(d+2)}},
\end{align}
where $\Sigma_{AB}^{min}$ is the minimal surface in the entanglement wedge $E_w(\rho_{AB})$ that ends on the minimal surface $\Gamma_{A\cup B}$, the dashed blue line in FIG. \ref{EWCS1}.
As a result, we have \cite{Takayanagi:2017knl, Nguyen:2017yqw}
\begin{align}\label{eop}
E_p(\rho_{AB})\equiv E_w(\rho_{AB}).
\end{align}
\section{Numerical result}\label{numerical}
\begin{figure}[]
\centering
\subfloat[$V_{\rm QCD}$ ]{\includegraphics[scale=0.33]{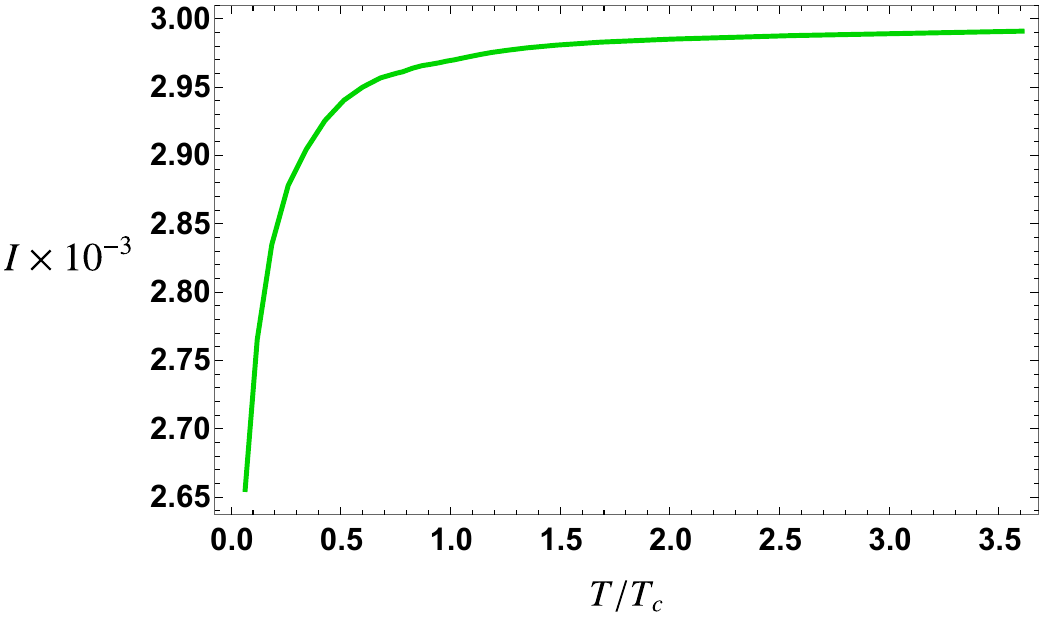} }
\subfloat[$V_{2{\rm nd}}$ ]{\includegraphics[scale=0.33]{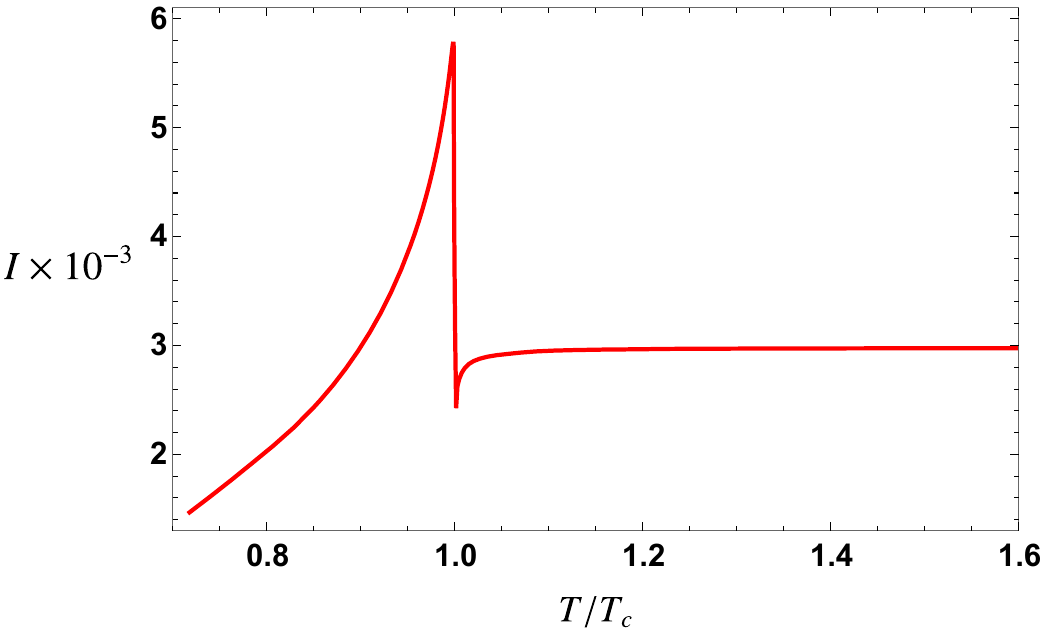}} 
\subfloat[$V_{1{\rm st}}$ ]{\includegraphics[scale=0.33]{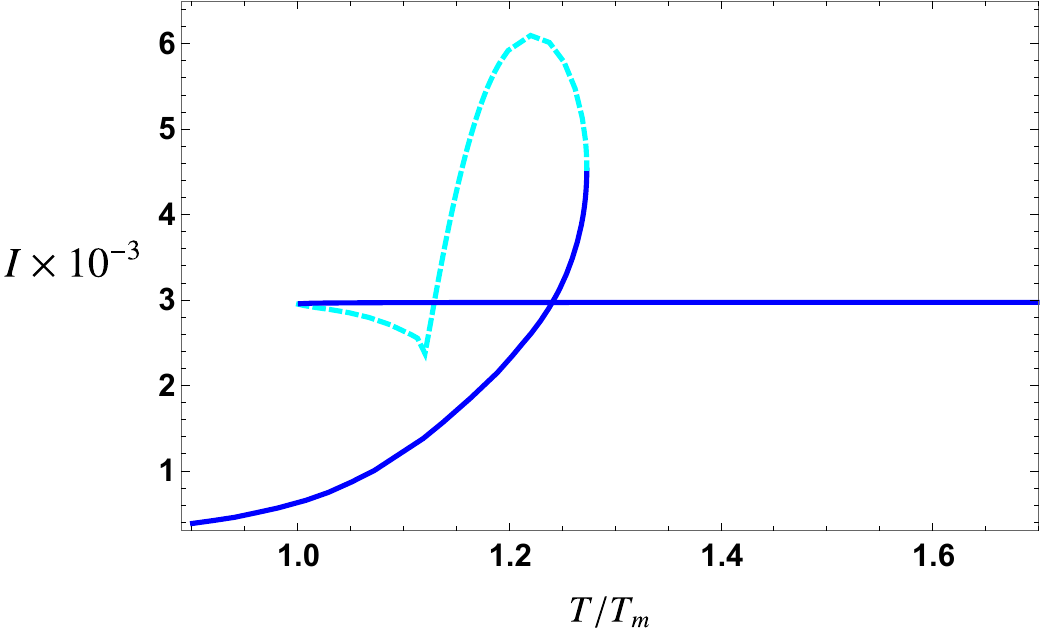}} \\
\subfloat[$V_{\rm QCD}$ ]{\includegraphics[scale=0.33]{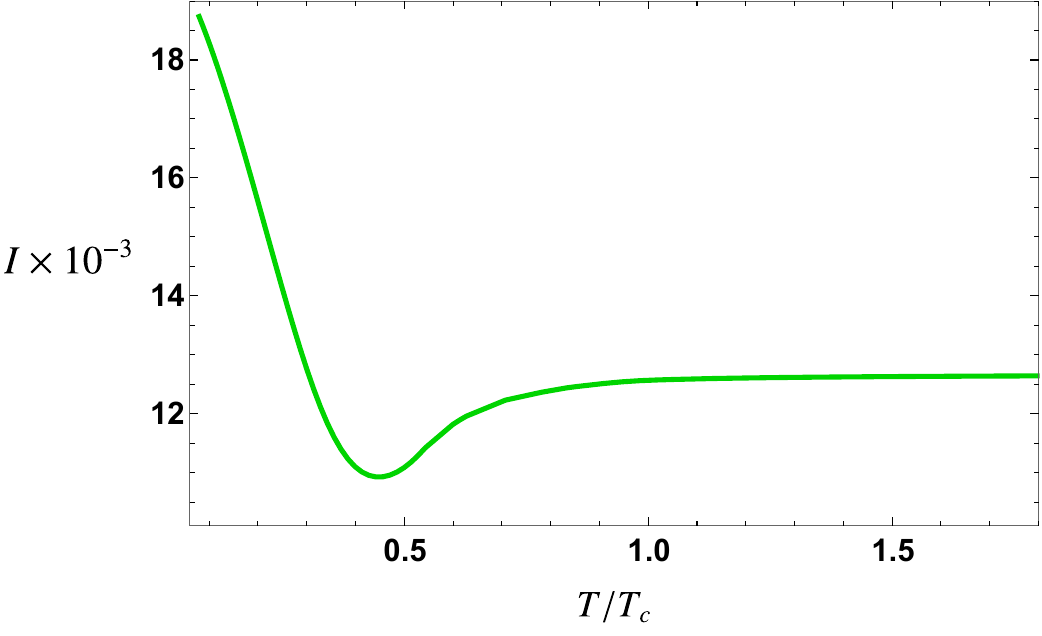} }
\subfloat[$V_{2{\rm nd}}$ ]{\includegraphics[scale=0.33]{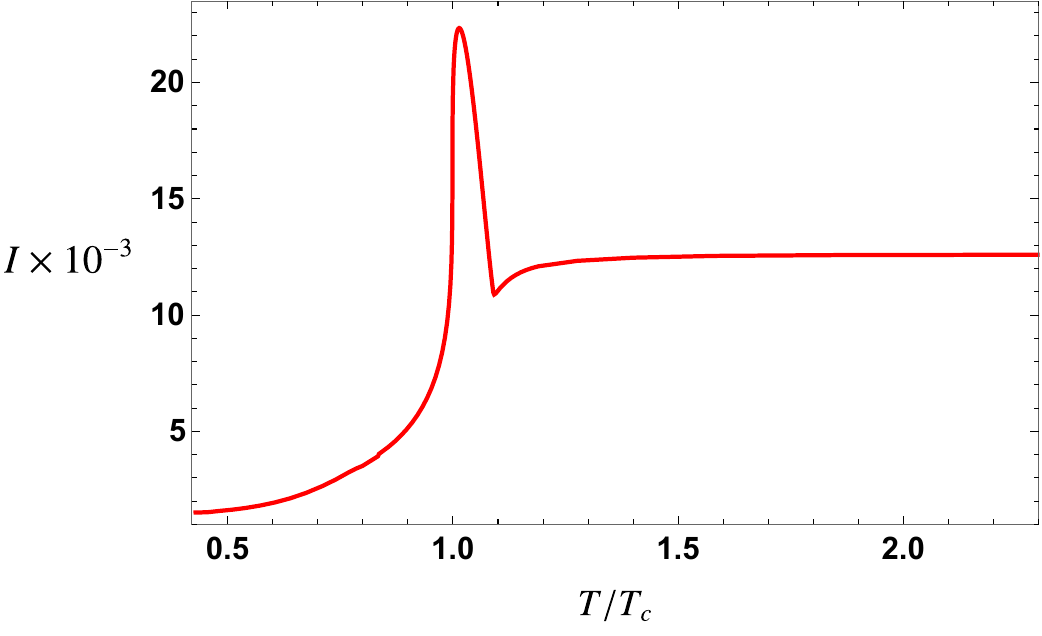}} 
\subfloat[$V_{1{\rm st}}$ ]{\includegraphics[scale=0.33]{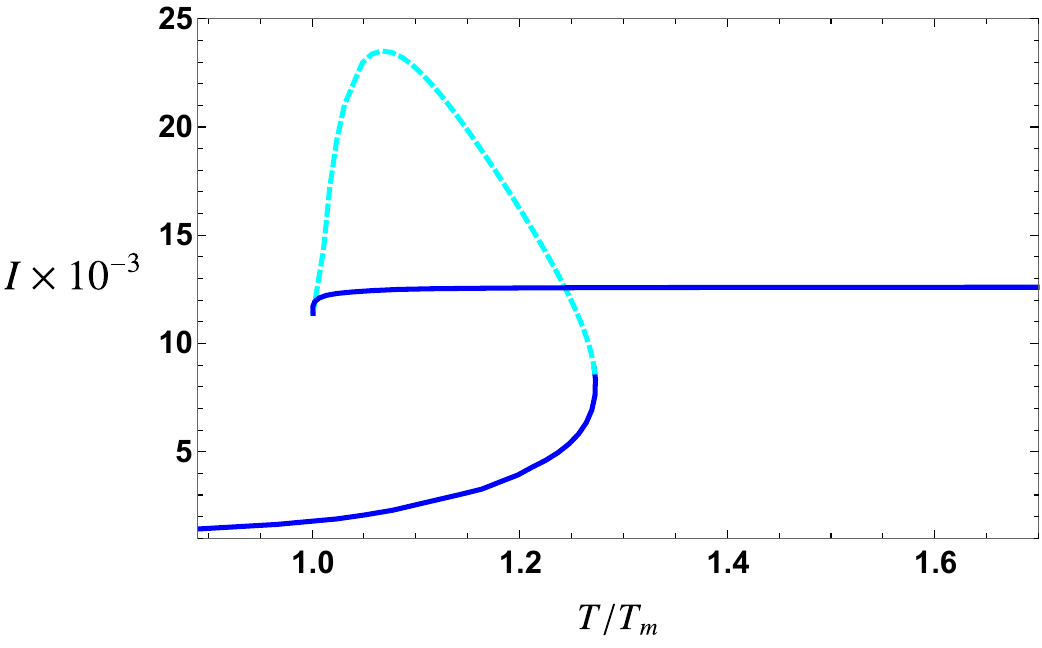}}
\captionsetup{justification=raggedright,singlelinecheck=false,font={small}}
\caption{Top row: The HMI  v.s. temperature $T$ for the dilaton potentials $V_{\rm QCD}$ (left), $V_{2{\rm nd}}$ (middle) and $V_{1{\rm st}}$ (right) for fixed $l=0.05$ and  $l'=0.01$. Bottom row: The HMI  v.s. temperature $T$ for the dilaton potentials $V_{\rm QCD}$ (left), $V_{2{\rm nd}}$ (middle) and $V_{1{\rm st}}$ (right) for fixed $l=0.05$ and  $l'=0.005$. In the right panels the solid blue curves correspond to the thermodynamically stable black hole solutions and the dashed cyan curves corresponds to the unstable solutions.}\label{EoPMI-over-T}
\end{figure}

\begin{figure}[]
\centering
\subfloat[$V_{\rm QCD}$ ]{\includegraphics[scale=0.33]{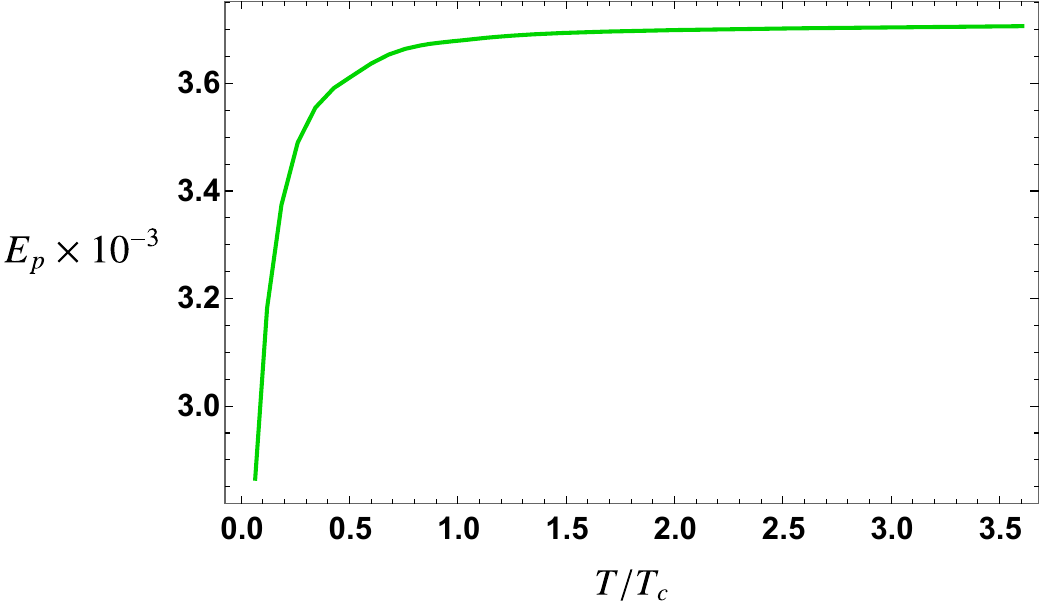}} 
\subfloat[$V_{2{\rm nd}}$]{\includegraphics[scale=0.33]{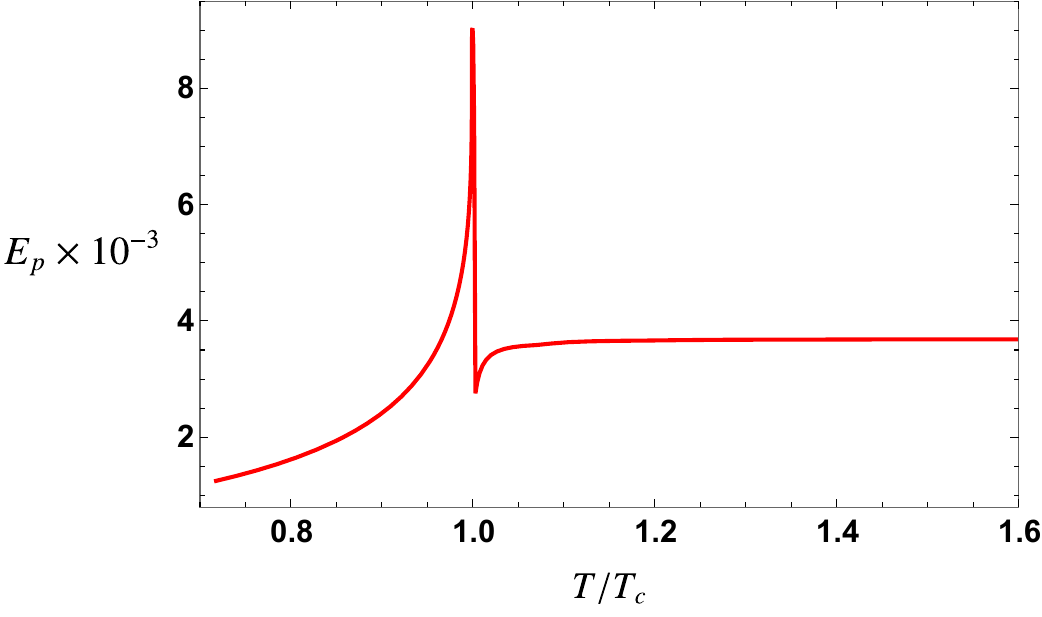}} 
\subfloat[$V_{1{\rm st}}$ ]{\includegraphics[scale=0.33]{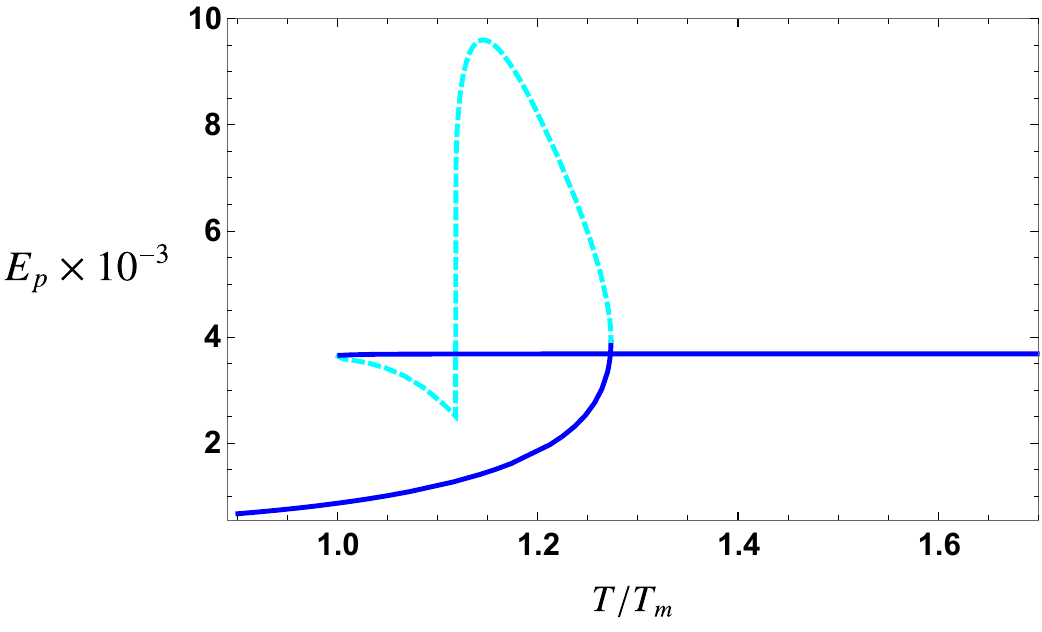}}\\
\subfloat[$V_{\rm QCD}$ ]{\includegraphics[scale=0.33]{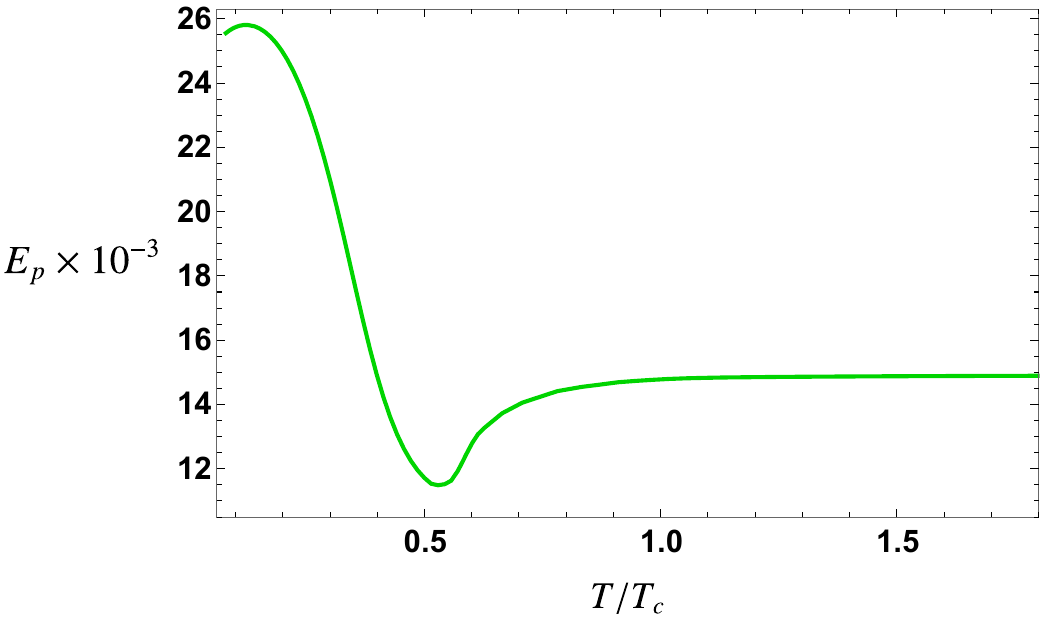}} 
\subfloat[$V_{2{\rm nd}}$]{\includegraphics[scale=0.33]{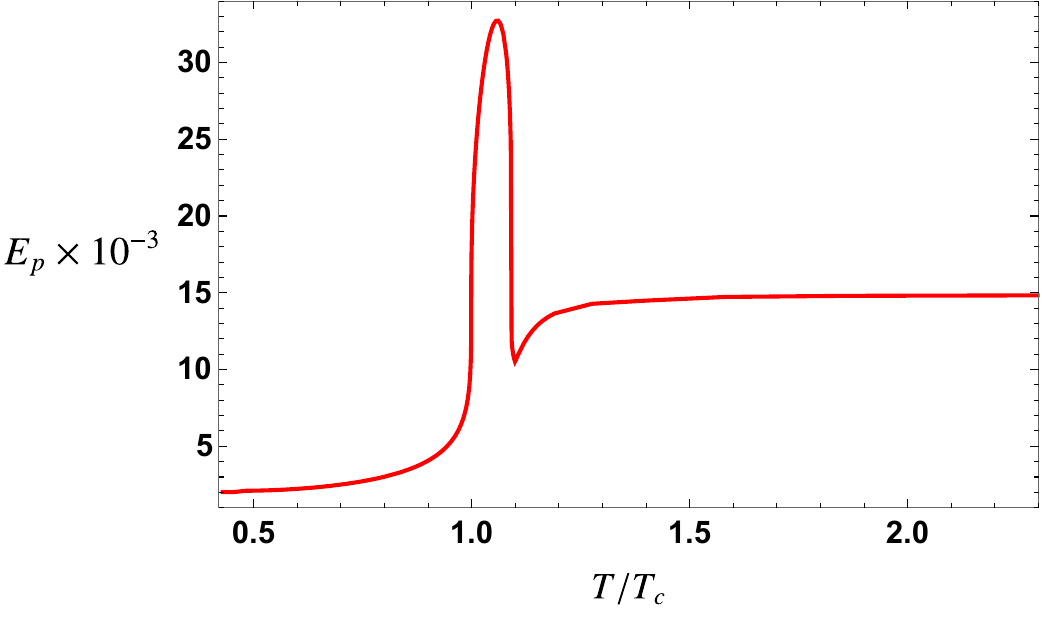}} 
\subfloat[$V_{1{\rm st}}$ ]{\includegraphics[scale=0.33]{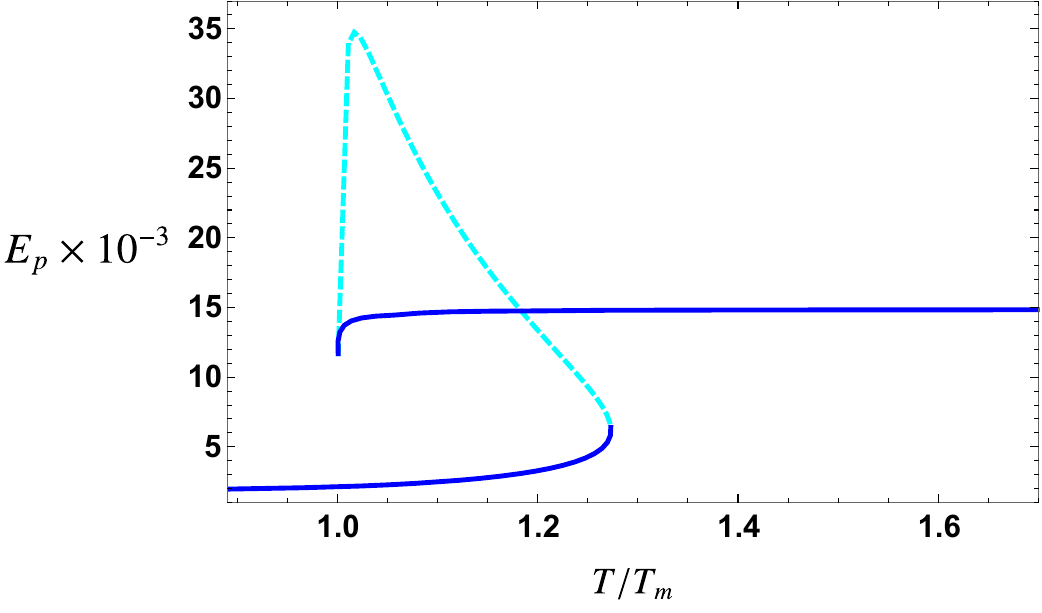}}
\captionsetup{justification=raggedright,singlelinecheck=false,font={small}}
\caption{Top row: The EoP  v.s. temperature $T$ for the dilaton potentials $V_{\rm QCD}$ (left), $V_{2{\rm nd}}$ (middle) and $V_{1{\rm st}}$ (right) for fixed $l=0.05$ and  $l'=0.01$. Bottom row: The EoP  v.s. temperature $T$ for the dilaton potentials $V_{\rm QCD}$ (left), $V_{2{\rm nd}}$ (middle) and $V_{1{\rm st}}$ (right) for fixed $l=0.05$ and  $l'=0.005$. In the right panels the solid blue curves correspond to the thermodynamically stable black hole solutions and the dashed cyan curves corresponds to the unstable solutions.}\label{EoPMI-over-T1}
\end{figure}

In order to study the HMI and EoP in the three cases of the dilaton potentials, we use \eqref{HMI} and \eqref{EWCS} and compute the area of the minimal surfaces and the entanglement wedge cross section for the metric \eqref{metric}. It is hard to do analytical calculation in the background \eqref{metric} and hence we use numerical methods to obtain the HMI and EoP. We compute the HMI and EoP for two different values of $l$ and $l'$ i.e. $(l=0.05, l'=0.01)$ and $(l=0.05, l'=0.005)$.
In FIGs. \ref{EoPMI-over-T} and \ref{EoPMI-over-T1}, we plot the HMI and EoP in terms of temperature for the dilaton potentials $V_{\rm QCD}$, $V_{2{\rm nd}}$ and $V_{1{\rm st}}$.
From these figures one can see that in the high temperature limit, the values of the HMI and EoP approaches to a constant values in the three cases of potentials which is equivalent to their conformal values. This behavior has been seen for the growth rate of holographic complexity density in \cite{Zhang:2017nth}. From the left panels in FIGs. \ref{EoPMI-over-T} and \ref{EoPMI-over-T1}, $V_{\rm QCD}$ case, we observe that the HMI and EoP have smooth behavior. In this case the HMI and EoP and their derivatives with respect to the temperature are continuous which show that there is a crossover phase transition at the critical temperature. The dependence of the HMI and EoP on temperature in the $V_{2{\rm nd}}$ case, middle panels, show that although at the critical temperature $T_c$ the HMI and EoP are both finite, their derivatives with respect to the temperature show a power law divergence in the vicinity of $T_c$. This indicates that the system possesses a second order phase transition at $T_c$. We will study the critical exponent using the HMI and EoP later on.  In the right panels of the FIGs. \ref{EoPMI-over-T} and \ref{EoPMI-over-T1}, we plot the HMI and EoP for the $V_{1{\rm st}}$ potential. Similar to the thermodynamic quantities in the range of temperature $T_m<T<1.272T_m$ which we mentioned in subsection \ref{thermo}, there are three branches of solutions which two of them correspond to the thermodynamically stable black hole solutions (solid blue curves) and one of them corresponds to the unstable black hole solution (dashed cyan curve). At the critical temperature $T_c\simeq 1.05T_m$ the HMI and EoP and their derivative with respect to the temperature are not continuous at $T_c$ which means the system possesses a first order phase transition at $T_c$.  In short, we conclude that the behavior of the HMI and EoP suggests that one can use these quantities to characterize  the type of phase transition. The interesting point is that although the shape of the HMI and EoP functions alter by changing $l$ and $l'$, their behavior does not change in the vicinity of $\frac{T}{T_c}=1$ and hence we can use them for probing the phase structure of the strongly coupled matter. This is in complete agreement with the result obtained for the second order phase transition in \cite{Amrahi:2020jqg}.
%\subsubsection{The crossover case}
%\subsubsection{The $2{\rm nd}$ order phase transition case}
\subsection{Critical exponent}
\begin{figure}[]
\centering
\includegraphics[scale=0.33]{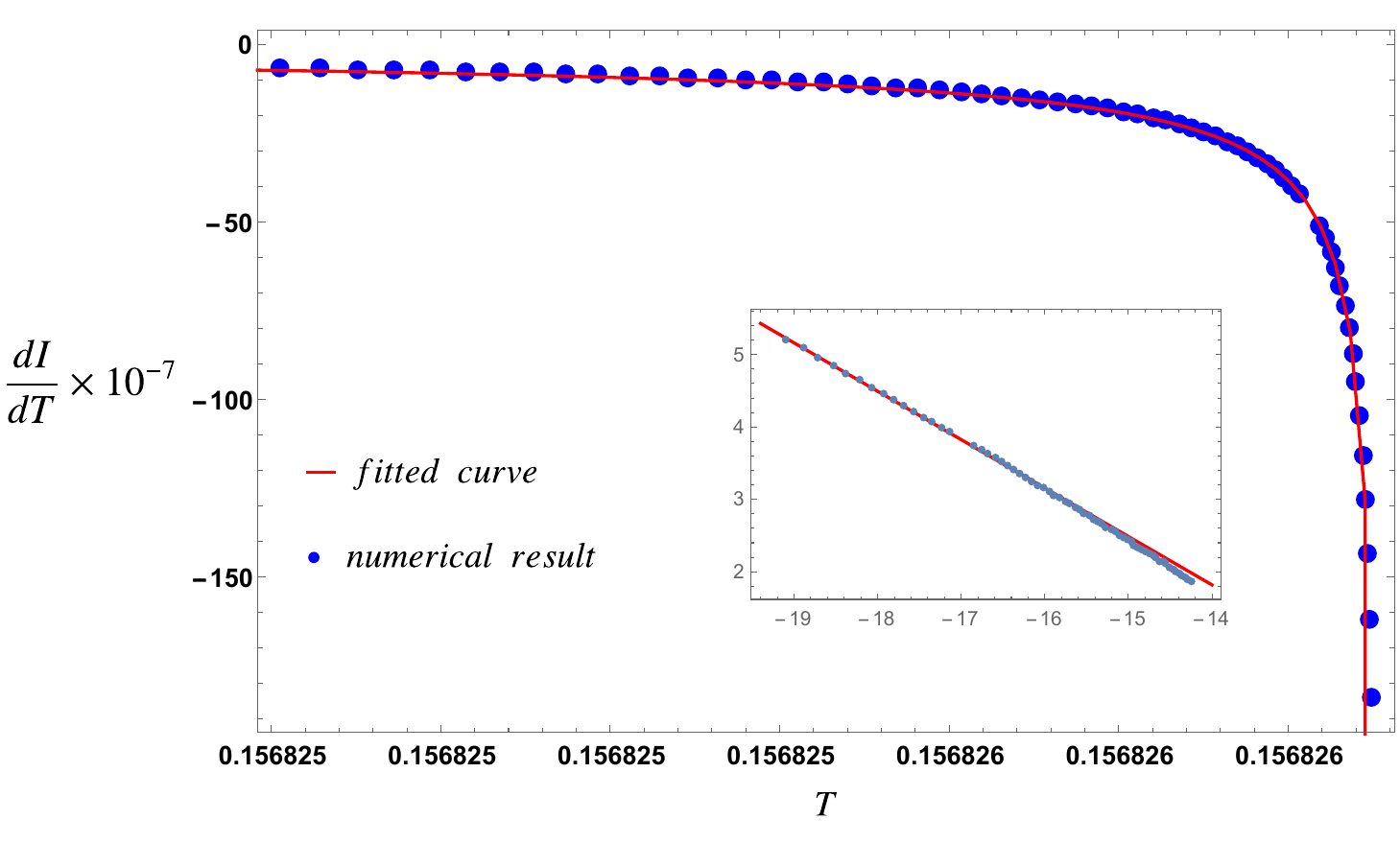} 
\includegraphics[scale=0.33]{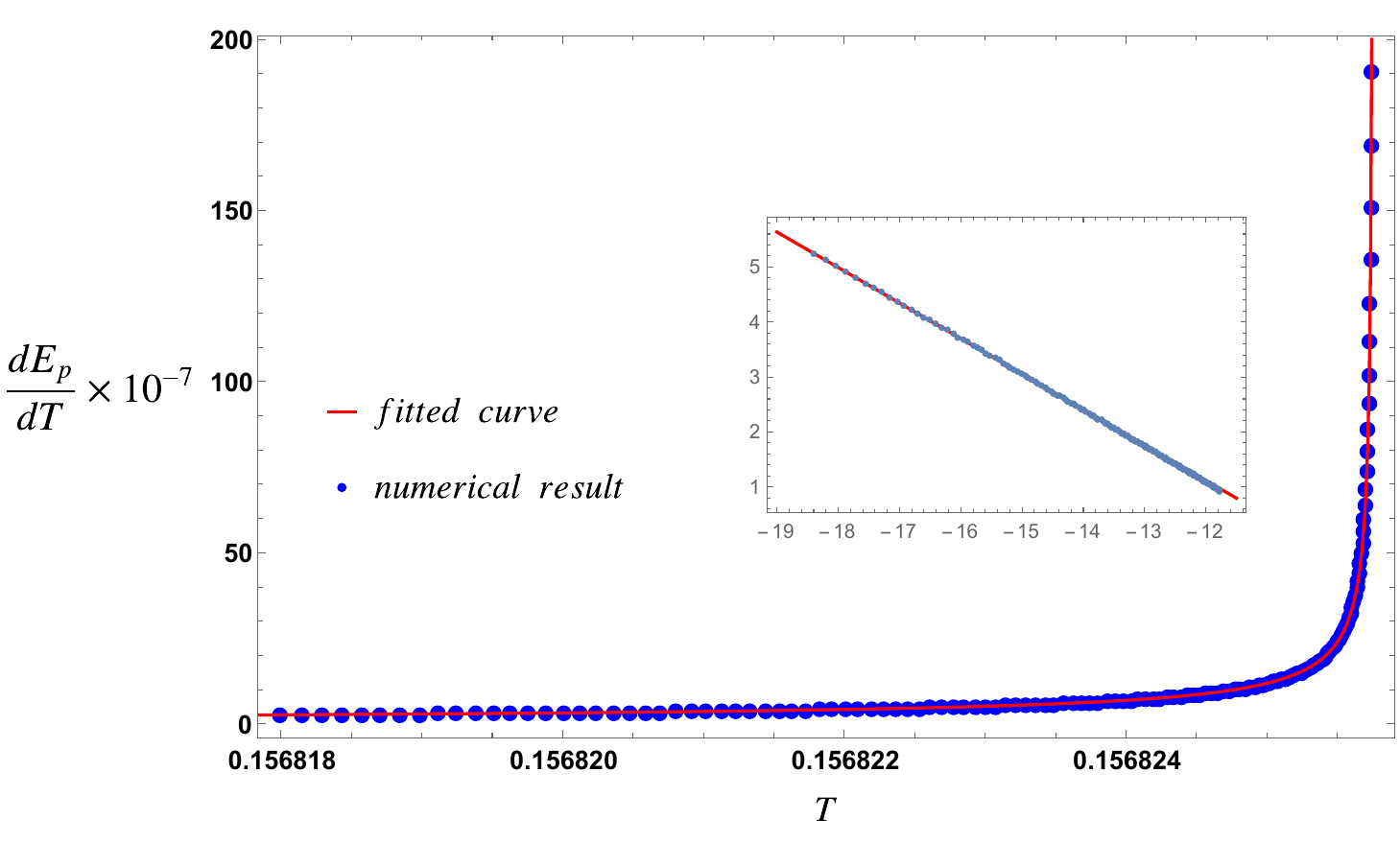} 
\captionsetup{justification=raggedright,singlelinecheck=false,font={small}}
\caption{Left: The slope of $I$ with respect to $T$ for fixed $l=0.05$ and  $l'=0.01$. The fitted curve with the numerical result is $\frac{dI}{dT}=-18289.51\ t^{-0.668}$. The corresponding RE and RMS are $0.003$ and $0.31$ respectively. The small plot is the logarithm of the data results and the linear function fitted with them.
Right: The slope of $E_p$ with respect to $T$ for fixed $l=0.05$ and  $l'=0.01$. The fitted curve with the numerical result is $\frac{dE_p}{dT}=39675.76 \ t^{-0.652}$. The corresponding RE and RMS are $0.021$ and $0.093$ respectively. The small plot is the logarithm of the data results and the linear function fitted with them.}\label{expo}
\end{figure}

When the system enjoys the second order phase transition at a critical temperature $T_c$, we saw that the HMI and EoP are continuous at $T_c$ while their slope with respect to temperature are not continuous. How the behavior of these quantities are near the critical temperature is an important question and we will discuss it in the following. We focus on the vicinity of the critical temperature in the HMI and EoP diagrams for $V_{2\rm nd}$ and this suggests that the slope of these plots for the near of $T_c$ can be fitted with a function of the form $t^{-\theta}$ where $t\equiv \frac{\vert T-T_c\vert}{T_c}$ and hence the number $\theta$, called the critical exponent, describes the variation of the HMI and EoP with respect to temperature. In FIG. \ref{expo}, we plot the slope of the HMI and EoP for $l=0.05$ and $l'=0.01$ near the critical temperature where we have defined the slope of a quantity $y$ with respect to temperature $T$ as
\begin{align}
\frac{dy}{dT}(i)=\frac{y(i+1)-y(i)}{T(i+1)-T(i)},
\end{align}
where $i$ represents the $i$th point in the corresponding data points. By fitting a curve with the numerical result, the value of the critical exponent is obtained $0.668$ and $0652$ from the HMI (left panel) and the EoP (right panel), respectively. These values are in agreement with one obtained from the behavior of the specific heat in \cite{Gubser:2008ny,Janik:2016btb}. We also plot the linear log-log diagrams for the HMI and EoP where the critical exponent is the slope of a line i.e. $\log (\frac{dy}{dT}) \propto \theta \log (t)$, and $y$ denotes the HMI or the EoP. We calculate the RE and RMS which are reported in the caption of FIG. \ref{expo}.
%\begin{figure}[]
%\centering
%\includegraphics[scale=0.45]{mutual-critical-exponent} 
%\caption{The slope of $I$ with respect to $T$. The fitted curve with the numerical result is $\frac{dI}{dT}=-5299.09\left(T_c-T\right)^{-0.668}$. The corresponding RE and RMS are $0.003$ and $0.31$ respectively. The small plot is the logarithm of the data results and the linear function fitted with them.}\label{MI-expo}
%\label{}
%\end{figure}
%\subsubsection{The $1{\rm st}$ order phase transition case}

\section{Conclusion}\label{conclution}
In this paper, we consider a non-conformal field theory at finite temperature which has holographic dual. This holographic model is used to mimic the equation of state of QCD by introducing a nontrivial dilaton field whose corresponding potential break the conformal symmetry. The dilaton potential are given by four parameters and the values of these parameters can be chosen to reproduce the lattice QCD results in which the system exhibits a crossover phase transition. Moreover, different choices of these parameters lead to different thermodynamical properties of this model. We choose three set of parameters for dilaton potential, labeled by $V_{\rm QCD}$, $V_{2{\rm nd}}$ and $V_{1{\rm st}}$ in which the system exhibits respectively the crossover, first and the second order phase transitions at a certain critical temperature. In this model, we study some thermodynamical quantities i.e. entropy density $s$, speed of sound $c_s^2$, specific heat $C_v$ and free energy $F$ whose behaviors confirm the phase structure of the system. We calculate the HMI and EoP for a symmetric configuration including two disjoint strip with equal width $l$ which are separated by the distance $l'$. Our result show that the behavior of the HMI and EoP can determine the type of the phase transition and hence we can use these entanglement measures to characterize the phase structures of strongly coupled matters. Moreover, at the second order phase transition, we focus on the near critical point and obtain the critical exponent using the EoP and HMI which is in agreement with the specific heat critical exponent.

Several problems call which we leave for further investigations. There are other quantum information quantities such as reflected entropy, odd entanglement entropy and logarithmic negativety and one can study them to probe various phase structures of the strongly coupled matters.  It would also be interesting to study the quantum information quantities in the holographic QCD model at finite temperature including chemical potential which can generate  more complete phase diagram of QCD \cite{DeWolfe:2010he}.

%\appendix
%\section{Some title}

%\acknowledgments
\section*{Acknowledgement}
We would like to kindly thank F. Taghinavaz for useful comments and discussions on related topics.

\end{document}